\documentclass[11pt,a4paper]{article}

\usepackage[a4paper,margin=1in]{geometry}
\usepackage[T1]{fontenc}
\usepackage[utf8]{inputenc}
\usepackage{newtxtext}
\usepackage{amsmath}
\usepackage{newtxmath}
\usepackage[final]{microtype}
\usepackage{setspace}
\usepackage{graphicx}
\graphicspath{{main/}{Supplemental-Information/}{images/}}
\usepackage{float}
\usepackage{subcaption}
\usepackage{adjustbox}
\usepackage{booktabs,array,makecell,tabularx}
\usepackage{multirow}
\usepackage[table]{xcolor}
\usepackage{ragged2e}
\usepackage{enumitem}
\usepackage{placeins}
\usepackage{caption}
\usepackage{rotating}
\usepackage{tikz}
\usepackage{docmute}
\usepackage{newunicodechar}
\usepackage{CJKutf8}

\usepackage[super,comma,sort&compress]{natbib}
\usepackage{bibunits}
\usepackage{xurl}

\definecolor{aoNavy}{HTML}{17324D}
\definecolor{aoBlue}{HTML}{0B5CAD}
\definecolor{aoLightBlue}{HTML}{EEF6FF}
\definecolor{aoRule}{HTML}{D7E8FA}
\definecolor{highlightcolor}{rgb}{0.88,1,1}
\definecolor{gray}{rgb}{0.9,0.9,0.9}
\definecolor{blue2}{rgb}{0.949,0.9686,0.9882}
\definecolor{badgeblue}{HTML}{30A7B3}

\usepackage[
  colorlinks=true,
  linkcolor=aoBlue,
  citecolor=aoBlue,
  urlcolor=aoBlue,
  pdfauthor={Yuliang Liu et al.},
  pdftitle={AlphaOracle: Oracle bone script decipherment via human-workflow-inspired deep learning}
]{hyperref}

\usepackage{titlesec}
\titleformat{\section}
  {\Large\bfseries\sffamily\color{aoNavy}}
  {\thesection.}{0.65em}{}
\titleformat{name=\section,numberless}
  {\Large\bfseries\sffamily\color{aoNavy}}
  {}{0pt}{}
\titleformat{\subsection}
  {\normalsize\bfseries\sffamily\color{aoNavy}}
  {\thesubsection}{0.6em}{}
\titleformat{name=\subsection,numberless}
  {\normalsize\bfseries\sffamily\color{aoNavy}}
  {}{0pt}{}
\titlespacing*{\section}{0pt}{2.2ex plus 0.4ex minus 0.2ex}{1.0ex}
\titlespacing*{\subsection}{0pt}{1.7ex plus 0.3ex minus 0.2ex}{0.7ex}

\defaultbibliographystyle{numbered}
\defaultbibliography{ref}

\newcommand{\CJKtrad}[1]{\begingroup\mdseries\begin{CJK*}{UTF8}{bsmi}#1\end{CJK*}\endgroup}
\newunicodechar{龝}{秋}

\newcommand{\AlphaInfoBox}[2]{%
  \par\medskip
  \noindent\begingroup
  \setlength{\fboxsep}{10pt}%
  \colorbox{aoLightBlue}{%
    \begin{minipage}{\dimexpr\linewidth-2\fboxsep\relax}
      {\sffamily\bfseries\color{aoNavy}#1}\par\vspace{0.35em}
      #2
    \end{minipage}%
  }%
  \endgroup\par\medskip
}

\newcommand{\AlphaPlainBox}[1]{%
  \par\smallskip
  \noindent\begingroup
  \setlength{\fboxsep}{10pt}%
  \colorbox{aoLightBlue}{%
    \begin{minipage}{\dimexpr\linewidth-2\fboxsep\relax}
      #1
    \end{minipage}%
  }%
  \endgroup\par\smallskip
}

\newcommand{\imgbadge}[1]{%
  \tikz[baseline=(char.base)]{
    \node[
      circle,
      fill=badgeblue,
      text=white,
      font=\sffamily\bfseries\small,
      minimum size=6.5mm,
      inner sep=0pt
    ] (char) {#1};
  }%
}

\newcommand{\imagewithbadge}[2]{%
  \begin{tikzpicture}
    \node[inner sep=0pt] (img) {\includegraphics[width=\linewidth]{#2}};
    \node[
      anchor=north west,
      xshift=0mm,
      yshift=9mm
    ] at (img.north west) {\imgbadge{#1}};
  \end{tikzpicture}%
}

\newsavebox{\mycaptionbox}
\DeclareCaptionFormat{singleleft}{%
    \sbox{\mycaptionbox}{#1#2#3}%
    \ifdim\wd\mycaptionbox<\linewidth
        \noindent\makebox[\linewidth][l]{#1#2#3}\par
    \else
        \noindent #1#2#3\par
    \fi
}

\newcommand{\frontmatterheading}[1]{%
  \par\vspace{0.45em}%
  \noindent{\large\bfseries\sffamily\color{aoNavy}#1}\par
  \vspace{0.2em}%
}

\title{AlphaOracle: Oracle bone script decipherment via human-workflow-inspired deep learning}
\date{}

\author{%
\begin{tabular}{c}
Yuliang Liu,\textsuperscript{1,\textdagger,*}
Haisu Guan,\textsuperscript{1,\textdagger}
Pengjie Wang,\textsuperscript{1,\textdagger}
Xinyu Wang,\textsuperscript{1,\textdagger}
Jinpeng Wan,\textsuperscript{1}
Kaile Zhang,\textsuperscript{1} \\
Handong Zheng,\textsuperscript{1}
Xingchen Liu,\textsuperscript{1}
Zhebin Kuang,\textsuperscript{1}
Huanxin Yang,\textsuperscript{1}
Bang Li,\textsuperscript{3}
Yongge Liu,\textsuperscript{3,*} \\
Lianwen Jin,\textsuperscript{2,*}
and Xiang Bai\textsuperscript{1,*}
\end{tabular}
}

\makeatletter
\newcommand{\makearxivtitle}{%
\begin{center}
  {\LARGE\bfseries\sffamily\color{aoNavy}\@title\par}
  \vspace{0.55em}
  {\small\@author\par}
  \vspace{0.55em}
  {\scriptsize
  \textsuperscript{1}School of Software Engineering, Huazhong University of Science and Technology, Wuhan 430074, China\\
  \textsuperscript{2}School of Electronic Information Engineering, South China University of Technology, Guangzhou 510641, China\\
  \textsuperscript{3}Key Laboratory of Oracle Bone Inscriptions Information Processing, Anyang Normal University, Anyang 455002, China\\
  \textsuperscript{\textdagger}These authors contributed equally\\
  \textsuperscript{*}Correspondence:
  \href{mailto:ylliu@hust.edu.cn}{ylliu@hust.edu.cn} (Y.L.);
  \href{mailto:liuyongge@aynu.edu.cn}{liuyongge@aynu.edu.cn} (Y.L.);\\
  \href{mailto:eelwjin@scut.edu.cn}{eelwjin@scut.edu.cn} (L.J.);
  \href{mailto:xbai@hust.edu.cn}{xbai@hust.edu.cn} (X.B.)\par}
  \vspace{0.45em}
  {\small\sffamily Accepted by \textit{The Innovation}. DOI: \href{https://doi.org/10.1016/j.xinn.2026.101462}{10.1016/j.xinn.2026.101462}\par}
\end{center}
\vspace{0.1em}
}
\makeatother

\newcommand*{\EnsureBibunitsAuxFile}{%
  \IfFileExists{bu.aux}{}{%
    \newwrite\alphaoraclebuaux
    \immediate\openout\alphaoraclebuaux=bu.aux
    \immediate\write\alphaoraclebuaux{\relax}%
    \immediate\closeout\alphaoraclebuaux
  }%
}

\newcommand*{\SupplementalBibliographyStyle}{Supplemental-Information/supplemental_numbered}
\newcommand*{\SupplementalBibliographyFile}{Supplemental-Information/supplemental_ref}

\begin{document}

\EnsureBibunitsAuxFile
\makearxivtitle

\frontmatterheading{GRAPHICAL ABSTRACT}

\begin{center}
    \includegraphics[width=\linewidth]{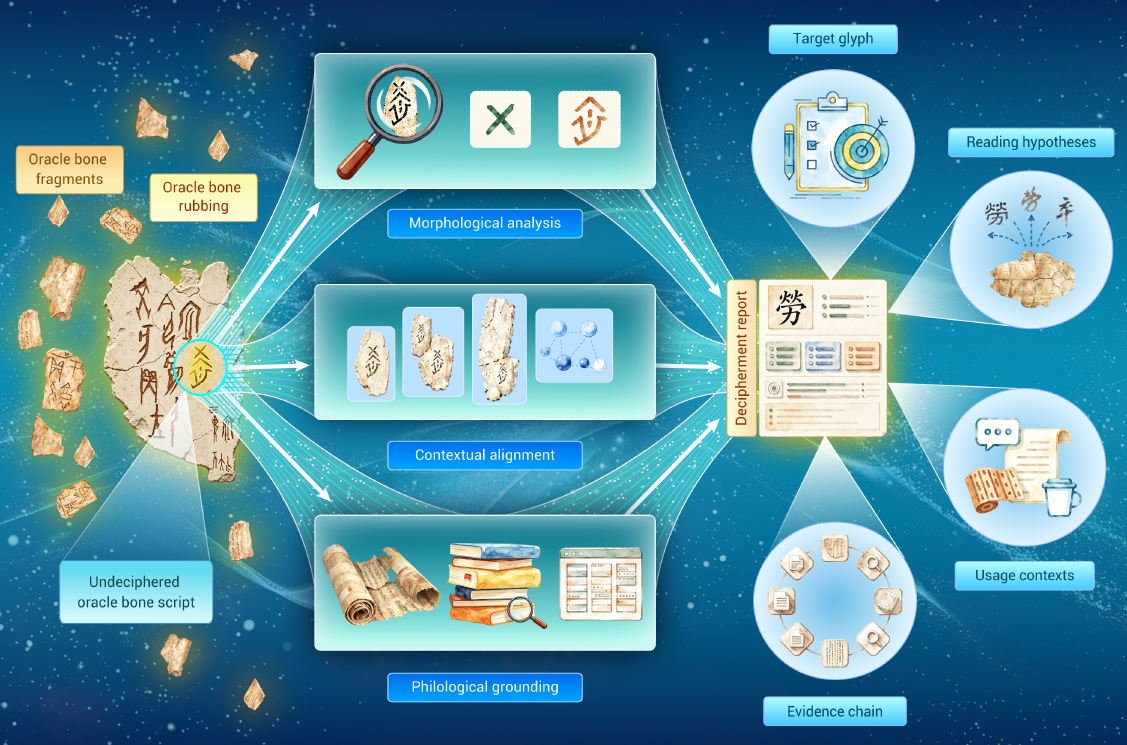}
\end{center}

\frontmatterheading{PUBLIC SUMMARY}

\AlphaPlainBox{%
\begin{itemize}[leftmargin=1.4em,itemsep=0.12em,topsep=0em]
    \item First AI system to emulate expert human workflow for oracle bone decipherment.
    \item Evidence-backed readings via morphology, context, and philology.
    \item Largest digitized corpus: 73,883 rubbings, 45,364 scans, 23,755 studies.
    \item Transparent, interpretable reports ready for expert review.
\end{itemize}
}

\clearpage

\AlphaInfoBox{Abstract}{%
Approximately 3,000 of the 4,500 oracle bone script (OBS) characters remain undeciphered due to fragmentary inscriptions and sparse evidence. Current AI approaches fail to replicate expert workflows that integrate form analysis, contextual semantics, and philological reasoning. We introduce AlphaOracle, a human-workflow-inspired framework that systematizes OBS decipherment using the largest digitized corpus to date. Its multi-stage pipeline comprises: (i) rubbing parsing; (ii) radical-based morphological analysis with diachronic modeling; (iii) contextual retrieval with semantic alignment; and (iv) philological validation against classical sources. Each stage generates explicit, confidence-weighted evidence chains, culminating in interpretable reports for scholarly verification. Across multiple test characters, AlphaOracle’s readings strongly agreed with expert interpretations. In a study of 86 domain specialists, it reduced analysis time by 64\% and 79\% of participants rated it highly useful. Notably, AlphaOracle resolves the character “\CJKtrad{勞}” as a toponymic or clan designation, offering concrete revisions to Shang administrative and social interpretations. These results suggest that computational methods aligned with philological practice can facilitate OBS research and provide a conceptual reference for studies of other undeciphered scripts.
}

\FloatBarrier
\vspace{0.5em}

\begin{bibunit}
\section{Introduction}

\begin{figure}[!ht]
    \centering
    \includegraphics[width=\linewidth]{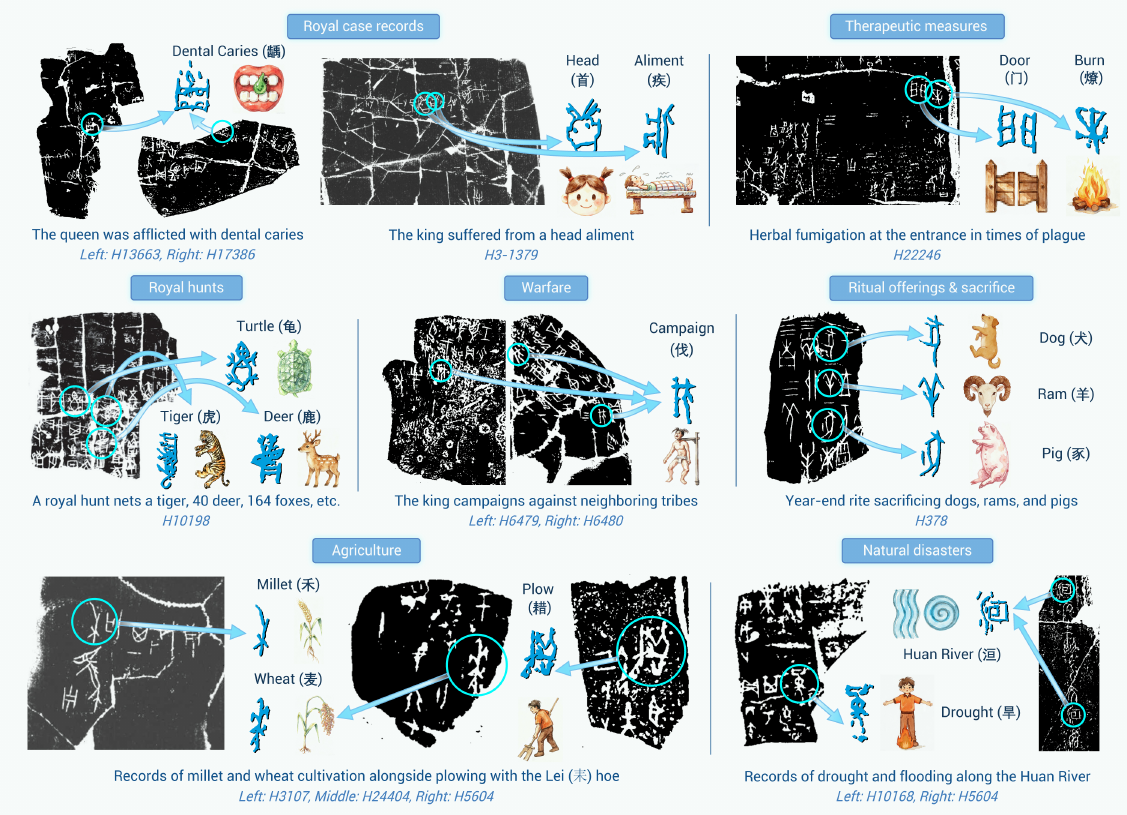}
    \caption{\textbf{Representative themes in oracle bone script} Oracle bone script documents a wide range of royal concerns during the Shang Dynasty, including health, therapeutic measures, warfare, ritual practice, agriculture, hunting, and natural disasters. The figure shows selected rubbings and corresponding characters across these domains, with inscription IDs (e.g., \textit{H13663}, \textit{H10168}) referring to cataloged oracle bone fragments.
    }
    \label{fig:OBI-Example}
\end{figure}

Oracle bone script (OBS), discovered in the late 19th century, represents the earliest known form of Chinese writing and a foundational component of East Asian epigraphy.\cite{chou1979chinese} Carved into ox scapulae and turtle plastrons during the Shang dynasty (circa 1250–1046 BCE),\cite{brunson2016new} these inscriptions document royal divination concerning hunting, health, agriculture, natural disasters, and ritual practice (see Figure~\ref{fig:OBI-Example}). To date, over 150,000 oracle bone fragments have been unearthed, bearing at least 4,500 distinct logographic symbols.\cite{li2021exploratory} Despite their historical and linguistic value, only about one-third of these scripts have been reliably deciphered, leaving much of this early writing system locked behind obscure characters, fragmentary texts, and lost contextual anchors.

In contrast to phonetic writing systems that rely on small, well-defined symbol inventories and transparent sound-script correspondences,\cite{assael2022restoring} OBS employs a logographic structure marked by extreme textual sparsity, orthographic variation, and the absence of phonetic annotations. These features hinder character recognition and also obscure higher-order semantic patterns, which complicates efforts to relate OBS to modern Chinese scripts. Further challenges arise from the physical state of the inscriptions, which are frequently fragmented, displaced from their original archaeological contexts, and stripped of syntactic regularity. These conditions make it difficult not only to identify known characters but also to interpret previously undeciphered ones, and pose further challenges when translating full oracle bone rubbings. To tackle these complexities, traditional epigraphic workflows rely on domain experts, whose interpretation of undeciphered characters typically involves analyzing historical variants, retrieving contextual and textual parallels from large reference corpora, and validating hypotheses across multiple inscriptions and transmitted sources. While research based on human efforts has produced significant insights,\cite{Jiang2018Bronze} it remains labor-intensive and difficult to scale due to its dependence on rare expert knowledge in character morphology, historical linguistics, and classical sources. Its inherent subjective, researcher-dependent variation and resistance to formal codification further limit reproducibility. These challenges emphasize the need for computational frameworks capable of systematically emulating expert reasoning across multiple stages of decipherment, while also scaling to analyze the vast bodies of existing data.

Deep learning has transformed fields as varied as virtual manuscript restoration or meteorological forecasting by extracting structure from vast,\cite{assael2022restoring,zhang2023skilful} noisy data. In the study of OBS, however, most computational efforts remain narrowly scoped: they primarily recognize and index characters already deciphered by Optical Character Recognition (OCR).\cite{liu2020oracle} While such systems accelerate preservation and retrieval, they leave the core problem largely untouched: deciphering the $\sim 3,000$ characters whose forms and meanings remain unknown.

Beyond transcription, the central challenge is to recover meanings for those undeciphered graphs through multi-stage reasoning. Traditional approaches have been effective for roughly 1,500 frequently attested characters, but they confront three persistent bottlenecks. First, \emph{extreme textual sparsity and fragmentation}: Many undeciphered forms occur infrequently, often in damaged contexts. Second, \emph{consensus evidence chains}: robust proposals must synthesize paleographic morphology, inscriptional context, and transmitted literature, the integration and assessment of such heterogeneous evidence often hinge on subjective expert judgments, making consensus difficult. Third, \emph{the difficulty of obtaining comprehensive validation}: evaluation is typically case-based and subjective, depending heavily on individual expertise rather than shared, comprehensive resources.

Recent studies have begun to push beyond simple transcription by applying computer vision models to propose modern equivalents from OBS inputs or by extending OCR-style pipelines to automate the recognition of known forms.\cite{guan2024deciphering,wang2024puzzle,liu2020oracle,lin2022radical} These methods, however, are largely \emph{purely visual}: they infer readings from shape alone and rarely marshal corroboration from contextual usage or transmitted historical texts. As a result, the outputs, although sometimes visually plausible, lack explanatory grounding, offer limited linguistic or historical justification, and provide only marginal assistance to domain experts. These challenges highlight the need for computational frameworks that not only emulate expert reasoning across multiple stages of decipherment but also \emph{scale to the analysis of large existing corpora}, ensuring that proposals are anchored in comprehensive, auditable evidence.

\begin{figure}[!t]
    \centering
    \includegraphics[width=1\linewidth]{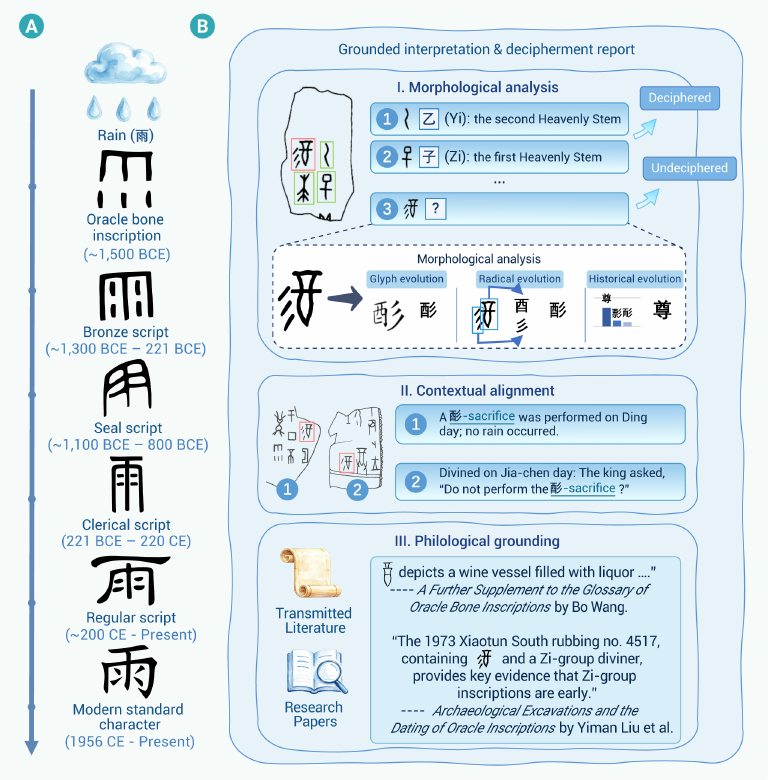}
    \caption{\textbf{Overview of character evolution and decipherment workflow} 
    (A) Evolution of the Chinese character `\CJKtrad{雨}' (Rain). 
    (B) Structure of the Decipherment Report generated by AlphaOracle, comprising \textit{morphological analysis}, \textit{contextual alignment}, and \textit{philological grounding}.}
    \label{fig:evolution}
    \vspace{-0.2cm}
\end{figure}

To bridge these gaps, we introduce \textbf{AlphaOracle}, a system designed to address the core challenges of OBS decipherment. AlphaOracle's aim is to recover the meanings of previously uninterpreted characters and to reconstruct inscriptions as coherent historical records grounded in rigorous, \textit{evidence-based} analysis. Unlike prior approaches that treat characters in isolation or focus narrowly on visual similarity, AlphaOracle emulates the multifaceted reasoning of human experts. It integrates established knowledge (secure readings, linguistic principles, accumulated scholarship, and transmitted historical literature) with inferences about unknown elements drawn from incomplete or unclassified inscriptions. As shown in Figure~\ref{fig:evolution}B, AlphaOracle assembles an evidence-backed argument for each proposed decipherment along three analytic dimensions: (i) \textit{Morphological Analysis}, which traces transformations across historical periods to infer function and predict modern equivalents for unknown glyphs; (ii) \textit{Contextual Alignment}, which retrieves occurrences from comprehensive digitized rubbings, models co-occurrence and usage patterns, and links ambiguous or rare symbols to context-specific meanings; and (iii) \textit{Philological Grounding}, which tests hypotheses against both ancient transmitted texts and modern scholarly research, situating them within the broader historical and linguistic record to ensure coherence beyond isolated readings. Empirical validation demonstrates that this integrated approach addresses longstanding challenges in OBS scholarship. For example, AlphaOracle has clarified contested readings such as `\CJKtrad{勞}’, resolved 22 disputed variant forms, and revealed new insights into Shang-period practices and interpretations. These achievements have been validated by domain experts who have judged the evidence chains to be credible and useful for scholarly adjudication, with several analyses incorporated into authoritative scholarly databases and receiving preliminary expert endorsement. 
By advancing AI-driven epigraphy from transcription toward integrated interpretation, AlphaOracle provides a conceptual framework that may inform studies of other undeciphered scripts and offers insights for digital humanities research and cultural heritage analysis.

\section{Materials and methods}

AlphaOracle addresses the fundamental challenges of oracle-bone decipherment through a systematic computational framework that emulates expert scholarly workflows while scaling to comprehensive corpus analysis. To bridge visual materials with the analytical workflow, an initial \textit{Rubbing Parsing Module} automatically processes oracle-bone rubbings, supplying inputs for the subsequent core analysis system. As illustrated in Figure~\ref{fig:evolution}B, the system integrates three complementary analytical dimensions to generate evidence-based decipherment reports for both individual characters and complete inscriptions. The framework operates through \textit{morphological analysis}, which decomposes unknown characters into constituent elements and traces their evolution across historical periods using a confidence-weighted ensemble of expert models. Iterative candidate interpretations from this stage are then evaluated through \textit{contextual alignment}, which combines corpus retrieval with sentence-level semantic modeling to assess interpretive plausibility within inscriptional contexts. Finally, \textit{philological grounding} tests these hypotheses against transmitted literature and modern scholarship through structured evidence retrieval, ensuring coherence with the broader historical and linguistic record.

Importantly, these components are not treated as three parallel decision-makers whose outputs are reconciled by a separate voting or meta-learning layer. Instead, AlphaOracle integrates evidence sequentially: \textit{morphological analysis} first produces a ranked set of candidate readings with calibrated confidence scores; \textit{contextual alignment} then re-scores these candidates for semantic plausibility while also providing complementary context-driven predictions; and \textit{philological grounding} subsequently evaluates the resulting candidate interpretations against external textual and glyphic evidence. For each input rubbing, AlphaOracle produces a final decipherment report that performs no additional adjudication but rather synthesizes pre-computed evidence across all three dimensions to present multiple candidate readings. This approach reflects the exploratory nature of scholarly interpretation while providing systematic, reproducible analysis at an unprecedented scale. Detailed architectures, training procedures, scoring functions, and implementation details are provided in the \hyperref[sec:supp_method]{Supplemental Methods} and Figure~\ref{fig:ex-networks}.

\subsection{Morphological analysis}
\label{sec:morphological}

The structure of oracle-bone characters provides a crucial key to their interpretation. Unlike alphabetic scripts, whose forms are largely arbitrary with respect to meaning, Chinese characters preserve a continuous visual lineage in which graphic form often encodes semantic content. As shown in Figure~\ref{fig:evolution}A, the character for \emph{“rain”} retains recognizable features across three millennia: cloud-like strokes above and falling droplets below, gradually stylized through bronze and seal scripts into its modern form. This continuity reflects two principles of Chinese morphology: meanings may persist through diachronic transformations, and complex characters are often built from meaningful subcomponents. AlphaOracle addresses these challenges through a confidence-weighted ensemble of three expert models trained on historical corpora spanning oracle bones, bronze inscriptions, transmitted texts, and modern lexicographical resources. The \textit{glyph evolution network} learns stylistic transformations, including simplification and standardization, to predict plausible modern equivalents for ancient forms; the \textit{radical evolution network} decomposes characters into radicals and models their combinatorial and evolutionary patterns, enabling inference even when complete characters are unattested; and the \textit{historical period evolution network} captures temporal variants and cross-period correspondences across dynasties and regions. Together, these models distinguish deciphered from unresolved instances and generate multiple candidate readings with calibrated confidence scores, providing uncertainty-aware inputs for downstream contextual and philological validation.

\subsection{Contextual alignment}
\label{sec:contextual}

While \textit{morphological analysis} proposes initial readings based on visual and structural evidence, these interpretations must be validated against inscriptional usage patterns. A glyph hypothesized to mean \emph{“rain”}, for instance, should appear in contexts related to weather, agriculture, or seasonal rites rather than only in military or administrative records. The \textit{contextual alignment} module formalizes this principle by retrieving candidate-glyph occurrences across the full corpus and analyzing their surrounding contexts with a BERT-based masked-language model trained on oracle-bone sentences. Users may also provide a hypothesized reading directly, prompting the system to retrieve all relevant instances and contextual references that would otherwise require extensive manual collation. Drawing on large-scale digitized oracle-bone collections, including 73{,}883 hand-traced rubbings and 45{,}364  archaeological scans, the module builds detailed usage profiles for each target glyph; the complete dataset sources, processing methods, and scales are summarized in Table~\ref{tab:alphaoracle-datasets}. By modeling co-occurrence patterns, distributional regularities, and formulaic structures characteristic of Shang divination texts, it evaluates how well each morphological candidate fits its textual environment. At inference, the module produces two complementary outputs: contextual re-ranking of candidates from \textit{morphological analysis} and independent semantic predictions over the full modern-character vocabulary. Beyond validating proposed readings, the same contextual evidence can support concise modern-language interpretations of complete inscriptions, offering crucial corroboration when visual or structural clues remain ambiguous.

\subsection{Philological grounding}
\label{sec:Philological}

Morphological and contextual analyses provide crucial evidence for character interpretation, but stronger validation comes from tracing proposed readings through transmitted textual traditions and accumulated scholarship. Although oracle bone script represents the earliest stage of Chinese writing, much of its core vocabulary, including terms for natural phenomena, implements, and ritual concepts, reappears in Western Zhou bronze inscriptions, Warring States manuscripts, Han lexica, and later classics. This continuity offers a third axis of validation: a plausible reading should leave semantic traces in later texts even as graphic forms evolve. The \textit{philological grounding} module systematizes this cross-textual analysis, which normally requires extensive expertise in classical Chinese and historical linguistics. It draws on a curated multi-tier corpus of 25 ancient transmitted texts and 23{,}755 authoritative modern scholarly studies, including the \textit{Shuowen Jiezi}, pre-Qin and Han classics such as the \textit{Yijing}, \textit{Shangshu}, \textit{Shijing}, \textit{Chunqiu}, and \textit{Zhuangzi}, as well as specialist references such as \textit{Guwenzi Gulin}, \textit{Jiaguwen Gulin}, and \textit{Jiaguwen Zilin}. Operationally, the module expands candidate queries with variant forms, graphic components, and near-synonyms, retrieves evidence through semantic text retrieval and glyph-image retrieval, and normalizes and deduplicates passages across editions. Retrieved evidence is weighted by relevance, source authority, genre, and historical period, allowing earlier and more reliable witnesses to contribute more strongly than later commentary. By linking graphic parallels, semantic correspondences, and diachronic continuity rather than relying on simple character matching, the module provides scholars with citable evidence bundles that support or challenge each proposed reading.

\section{Results}

Evaluating AlphaOracle requires protocols that capture both technical performance and practical scholarly utility. Due to the absence of established benchmarks for oracle bone decipherment, we design comprehensive evaluation protocols spanning computational metrics and expert assessment to validate the system's effectiveness across multiple dimensions.

\subsection{Evaluation setting}
We adopt a two-tier evaluation strategy addressing distinct aspects of the decipherment challenge: quantitative evaluation employs controlled datasets and computational metrics to assess individual component performance and end-to-end system capabilities, following standard practices in computational linguistics and computer vision, while epigrapher evaluation engages domain specialists in oracle bone studies to assess the scholarly utility, interpretive coherence, and practical value of system outputs through structured human studies.

For quantitative evaluation, we construct evaluation datasets by partitioning available oracle bone materials into training, validation, and test sets, ensuring no character overlap between splits. Each component of AlphaOracle is assessed on tasks that directly reflect its intended function within the decipherment workflow. \textit{Morphological analysis} is evaluated through character identification accuracy using a simulated decipherment scenario: we withhold known characters from the training data and test whether the system can correctly identify them based solely on visual and structural evidence, mimicking the challenge of interpreting truly unknown glyphs. \textit{contextual alignment} is assessed using two complementary approaches that measure the system's ability to leverage surrounding text for interpretation. The first approach systematically removes individual characters from the complete oracle bone script and evaluates whether the system can predict the missing character from context alone, analogous to how scholars often infer damaged or unclear characters by analyzing the surrounding text and formulaic patterns in divination records. The second approach presents the system with multiple candidate readings for a character (generated by \textit{morphological analysis}) and tests whether it can select the most contextually appropriate interpretation based on co-occurrence patterns and semantic coherence within the inscription. Interpretation quality is measured by comparing the system's modern-language translations of oracle bone texts against expert-curated references using established metrics that assess lexical accuracy, structural similarity, and semantic fidelity. Due to the specialized nature of oracle bone decipherment, we establish baselines using strong general language models (GPT-5, DeepSeek, Gemini 2.5 Pro) and domain-adapted systems where available, acknowledging that no existing systems are specifically designed for this task.

\subsection{Quantitative evaluation}

Table~\ref{tab:component-evaluation} presents quantitative results across three core evaluation dimensions, comparing AlphaOracle against available baselines on standardized tasks. We evaluate the \textit{morphological analysis} ensemble on a simulated decipherment task using 88 withheld characters from our annotated corpus. Under identical training conditions, AlphaOracle achieved a top-1 accuracy of 23.1\%, markedly outperforming existing image generation approaches, including our reimplementation of Google’s SR3 model. This performance demonstrates the effectiveness of integrating structural, evolutionary, and compositional evidence rather than relying on single analytical approaches. Two complementary tasks assess the contribution of distributional semantics to character interpretation through \textit{contextual alignment}. In the recovery task, which systematically restores characters from incomplete inscriptions, AlphaOracle achieves 56.1\% accuracy compared to 22.2\% for the best LLM baseline. The reranking task, which selects optimal readings from morphological candidates based on contextual fit, demonstrates even stronger performance at 95.2\% accuracy versus 65.4\% for the best language model baseline. These results validate the critical role of context in resolving interpretive ambiguities that remain intractable from morphological evidence alone. 

\begin{table}[htbp!]
\caption{Core performance metrics of AlphaOracle}

\centering
\renewcommand{\arraystretch}{1.3}
\setlength{\tabcolsep}{4pt}
\begin{adjustbox}{max width=\textwidth}
\begin{tabular}{c|cc|ccc|cccc}
\toprule
\multirow{5}{*}[-1em]{\rotatebox[origin=c]{90}{Component}} 
& \multicolumn{2}{c|}{\cellcolor{gray} \textbf{Glyph Analysis}} 
& \multicolumn{3}{c|}{\cellcolor{gray} \textbf{Contextual Restoration}} 
& \multicolumn{4}{c}{\cellcolor{gray} \textbf{Interpretation}} 

\\
& \textbf{Method} 
& \textbf{ACC} 
& \textbf{Method} 
& \textbf{Recovery} 
& \textbf{Reranking} 
& \textbf{Method} 
& \textbf{BLEU} 
& \textbf{ROUGE} 
& \textbf{METEOR} 

\\
\cmidrule{2-10}
\addlinespace[-4pt]
& Pix2Pix & NAN & GuwenBERT & 0.074 & 0.472 & GPT-5 & 0.133 & 0.195 & 0.285 \\

& CycleGAN & NAN & GPT-5 & 0.106 & 0.571 & Gemini 2.5 Pro & 0.163 & 0.225 & 0.337 \\
& SR3 & 0.131 & Gemini 2.5 Pro & 0.222 & 0.654 & DeepSeek & 0.171 & 0.229 & 0.346 \\

& \cellcolor{blue2} AlphaOracle & \cellcolor{blue2} \textbf{0.231} & \cellcolor{blue2} AlphaOracle & \cellcolor{blue2} \textbf{0.561} & \cellcolor{blue2} \textbf{0.952} & \cellcolor{blue2} AlphaOracle & \cellcolor{blue2} \textbf{0.491} & \cellcolor{blue2} \textbf{0.586} & \cellcolor{blue2} \textbf{0.659}\\

\end{tabular}
\end{adjustbox}

\setlength{\tabcolsep}{4pt}
\begin{adjustbox}{max width=\textwidth}

\begin{tabular}{c|ccc|cccc|cc}
\toprule
\multirow{3}{*}[-0.6em]{\rotatebox[origin=c]{90}{Human}} 
& \multicolumn{3}{c|}{\cellcolor{gray} \textbf{Overall Metrics}} 
& \multicolumn{4}{c|}{\cellcolor{gray} \textbf{Component Metrics}}  
& \multicolumn{2}{c}{\cellcolor{gray} \textbf{Retrieval Metrics}} 
\\
& \textbf{\makecell{No. of \\experts}} 
& \textbf{\makecell{High \\approval}} 
& \textbf{\makecell{Efficiency\\ gain}} 
& \textbf{\makecell{Rubbing \\ parsing}} 
& \textbf{\makecell{Morphological\\ analysis}} 
& \textbf{\makecell{Contextual \\alignment}} 
& \textbf{\makecell{Philological \\grounding}} 
& \textbf{Hit Rate}
& \textbf{ACC}
\\
\cmidrule{2-10}
\addlinespace[-2pt]

& \cellcolor{blue2} 86 & \cellcolor{blue2} 78.7\% & \cellcolor{blue2} 64.0\% & 
\cellcolor{blue2} 4.20 & \cellcolor{blue2} 4.44 & \cellcolor{blue2} 4.38  & \cellcolor{blue2} 4.01
 & \cellcolor{blue2} 90.2\% & \cellcolor{blue2} 74.8\%\\
\bottomrule
\end{tabular}
\end{adjustbox}
\caption*{The upper part reports objective evaluations across glyph analysis, contextual restoration, and interpretation tasks. The lower part presents expert evaluation on overall metrics, component metrics (including \textit{rubbing parsing}, \textit{morphological analysis}, \textit{contextual alignment}, and \textit{philological grounding}), and retrieval metrics.}
\label{tab:component-evaluation}
\end{table}

To evaluate the interpretation component, we assess the system's ability to generate coherent modern-language explanations of oracle bone texts, using this translation-like task as a proxy for measuring how well the system integrates morphological and contextual evidence into meaningful interpretations. We compare AlphaOracle's outputs against expert-curated references using established metrics that measure lexical accuracy (BLEU, how many words match), structural similarity (ROUGE, how well the overall meaning is captured), and semantic fidelity (METEOR, how closely the interpretation aligns with expert understanding). AlphaOracle demonstrates substantially higher performance across all measures: achieving scores of 0.491 versus 0.171 for lexical accuracy, 0.586 versus 0.229 for structural similarity, and 0.659 versus 0.346 for semantic fidelity compared to the best general language model baselines. It should be noted that these general models were evaluated in a zero-shot setting without domain-specific fine-tuning. The stark performance gap demonstrates that even state-of-the-art general reasoning models struggle with this highly specialized task out of the box, thereby validating our newly constructed OBS dataset as an essential and irreplaceable foundation for paleographic decipherment. These results indicate that effective oracle bone interpretation requires domain-specific expertise and systematic integration of paleographic, contextual, and philological evidence that even state-of-the-art language models cannot achieve without specialized and comprehensive historical knowledge integration. To further examine whether the interpretation component depends on a specific pre-trained language model, we additionally replaced the default Qwen2.5-7B backbone with two English-centric instruction-tuned models, Llama3.1-8B-Instruct and Ministral-8B-Instruct, while keeping the remaining training and evaluation protocol unchanged. As shown in Table~\ref{tab:interpretation}, both alternative backbones achieved competitive performance on the same interpretation task, although Qwen2.5-7B remained the strongest overall. These results suggest that the proposed framework is not tied to a single Chinese-centric language model, while also confirming that a Chinese-centric backbone is advantageous for generating modern Chinese explanations of oracle bone inscriptions.

\subsection{Expert assessment}

As no prior system provides end-to-end human-workflow decipherment for direct comparison, we adopted expert assessment as the primary evaluation method. A total of 86 specialists and doctoral researchers across multiple domains participated anonymously. To ensure transparency and reproducibility while maintaining participant anonymity, the detailed demographic distribution of the participating experts and the comprehensive breakdown of their questionnaire responses are provided in Table~\ref{tab:evaluation_metrics}. Each evaluation included the original rubbing, AlphaOracle's candidate interpretations, and complete evidence chains spanning \textit{morphological analysis}, \textit{contextual alignment}, and \textit{philological grounding}. A separate blind Human Arena evaluation further compared anonymized interpretations from AlphaOracle and baseline models, with the interface and results shown in Figure~\ref{fig:arena_interface} and Figure~\ref{fig:arena_results}.

Experts unanimously agree that AlphaOracle is helpful for decipherment tasks, with 78.7\% rating its usefulness as 4 or 5 (where 5 means extremely helpful, 4 means helpful, 3 means possibly helpful, 2 means slightly helpful, and 1 means not helpful at all). This indicates a high overall level of credibility and suggests that AlphaOracle reduces the required time by 64.0\%. In addition to these overall metrics, experts also provided direct usefulness ratings for each of the four major components. Expert assessments showed uniformly high endorsement across modules, with average usefulness scores of 4.20 for \textit{rubbing parsing}, 4.44 for \textit{morphological analysis}, 4.38 for \textit{contextual alignment}, and 4.01 for \textit{philological grounding} (Table~\ref{tab:component-evaluation}). Manual assembly of cross-textual parallels across extensive philological sources is time-consuming. In our \textit{philological grounding} evaluation, we randomly sampled 42 oracle bone characters and retrieved relevant materials from the corpora listed in Table~\ref{tab:alphaoracle-datasets}, leveraging both glyphic and semantic evidence. Because gold-standard retrieval annotations are unavailable for such a large corpus, we used sampled human evaluation to estimate performance, yielding a 90.2\% hit rate and 74.8\% precision. These results indicate that AlphaOracle attains high retrieval accuracy in substantially less time than traditional workflows, thereby reducing the manual burden in philological research.

Part of our analytical findings have been recognized by paleographers, as evidenced by their incorporation into an authoritative database. Notable examples include the long-debated character ``\CJKtrad{勞}'', for which scholarly opinion has been divided between two readings: one identifying it as a clan or place name, and another interpreting it as an adverb marking the completion of an event. AlphaOracle’s integrated evidence chain, built through \textit{morphological analysis}, distributional contexts within divinations, and philological validation from transmitted texts, corroborates the former view and supports its interpretation as a toponym or clan designation rather than an adverbial marker. This case illustrates how the system augments expert judgment by consolidating dispersed forms of evidence and making reasoning transparent, thereby enabling more efficient evaluation of contested readings and contributing to a cumulative scholarly consensus.

\section{Discussion}

Oracle bone decipherment remains one of archaeology’s enduring challenges, with roughly two-thirds of characters still without secure readings. AlphaOracle addresses this gap with a systematic framework that emulates expert practice by integrating \textit{morphological analysis}, \textit{contextual alignment}, and \textit{philological grounding}. The pipeline improves accuracy and efficiency while preserving explicit evidence chains and uncertainty, helping convert fragmentary inscriptions into interpretable historical narratives.

More broadly, this study provides a practical reference for ancient-script research scenarios in which the available evidence is fragmented or unorganized, but still accompanied by a meaningful body of related scholarly materials. It suggests that \textit{morphological analysis}, \textit{contextual alignment}, and \textit{philological grounding} can be combined within a computational framework informed by expert workflows, thereby helping structure evidence and support interpretation. At the same time, the approach remains limited for scripts or symbols that lack sufficient historical context, transmitted traditions, or supporting literature. We therefore view this work not as a general solution for all undeciphered writing systems, but as a modest step that may inform AI-assisted research on partially documented ancient inscriptions.

\section*{Resource availability}
\subsection*{Materials availability}
This study did not generate new unique materials/reagents.

\subsection*{Data and code availability}
\paragraph{\textbf{\textit{Data availability.}}} To train the model described in this study, we compiled and integrated a diverse corpus encompassing oracle bone rubbings and transcriptions, open-source repositories, ancient classical texts, scholarly journal articles, and authoritative monographs related to oracle bones. The specific utilization and allocation of these datasets are detailed within the ``Training details'' section, accessible via \url{http://vlrlabmonkey.xyz:7685/wenxian?lan=en}, with a live demo showing the end-to-end pipeline (\textit{rubbing parsing} $\rightarrow$ \textit{morphological analysis} $\rightarrow$ \textit{contextual alignment} $\rightarrow$ \textit{philological grounding} $\rightarrow$ interpretation). Comprehensive descriptions of the data content are presented in Table~\ref{tab:alphaoracle-datasets}. The oracle bone rubbing images used in this study were collected from publicly available academic resources, including published books, journal articles, open databases, and institutionally accessible digital collections. Our use of these materials is strictly for non-commercial scientific research and follows applicable copyright regulations.

\paragraph{\textbf{\textit{Code availability.}}} All source code, pretrained checkpoints, and data-preprocessing scripts for AlphaOracle are openly available at \url{https://github.com/Yuliang-Liu/AlphaOracle}. 

\section*{Funding and acknowledgments}
This work was supported by the National Natural Science Foundation of China (62225603, 6257614). The funders had no role in study design, data collection and analysis, decision to publish, or preparation of the manuscript. We thank thirty-one oracle-bone script scholars, epigraphers, digital-humanities researchers, museum professionals, independent scholars, and cultural-heritage practitioners for reviewing the system and for offering valuable feedback, suggestions, and expert support. Their expertise spans major universities, museums, research institutes, and oracle-bone research laboratories, including the Palace Museum, Henan University, Jilin University, Nanjing University, Tsinghua University, Capital Normal University, Fudan University, Shanghai Museum, Zhengzhou University, the Chinese Academy of Social Sciences, Henan Normal University, Ocean University of China, Southwest University, and the Oracle Bone Inscriptions Information Processing Laboratory, together with related scholarly communities in Korea, Japan, and the United States.

\section*{Author contributions}
Yuliang Liu, Haisu Guan, Pengjie Wang, and Xinyu Wang conceived the study. 
Haisu Guan and Jinpeng Wan developed the \textit{rubbing parsing} module.
Huanxin Yang developed a transformation model from rubbings to transcriptions. 
Pengjie Wang, Haisu Guan, and Jinpeng Wan developed the \textit{morphological analysis} module. 
Haisu Guan, Handong Zheng, and Zhebin Kuang developed the \textit{contextual alignment} module. 
Pengjie Wang and Xingchen Liu performed text digitization, and together with Handong Zheng developed the \textit{philological grounding} module.
Kaile Zhang and Jinpeng Wan developed the front end, and all authors contributed to the back end.
Zhebin Kuang curated literature and processed data. 
Yuliang Liu and Xinyu Wang contributed to methodology discussions and manuscript structure. 
Yuliang Liu and Yongge Liu coordinated communications with experts.
Bang Li and Yongge Liu provided oracle bone data and domain expertise. 
Lianwen Jin and Xiang Bai supervised the study. 
Xiang Bai provided funding support. 
All authors contributed to the manuscript and approved the final version.

\section*{Declaration of interests}
The authors declare no conflicts of interest.

\putbib

\end{bibunit}

\clearpage
\appendix
\setcounter{section}{0}
\renewcommand{\thesection}{Appendix~\Alph{section}}
\renewcommand{\theHsection}{appendix.\Alph{section}}
\titleformat{\section}
  {\Large\bfseries\sffamily\color{aoNavy}}
  {\thesection.}{0.65em}{}
\defaultbibliographystyle{\SupplementalBibliographyStyle}
\defaultbibliography{\SupplementalBibliographyFile}

\section*{\centering\LARGE Supplemental Information}
\begin{bibunit}

\setcounter{figure}{0}
\setcounter{table}{0}
\renewcommand{\thefigure}{S\arabic{figure}}
\renewcommand{\thetable}{S\arabic{table}}

\renewcommand{\theHfigure}{S\arabic{figure}}
\renewcommand{\theHtable}{S\arabic{table}}

\textbf{AlphaOracle: Oracle bone script decipherment via human-workflow-inspired deep learning}
\paragraph{DOI:} https://doi.org/10.1016/j.xinn.2026.101462
\paragraph{Authors List:} Yuliang Liu, Haisu Guan, Pengjie Wang, Xinyu Wang, Jinpeng Wan, Kaile Zhang, Handong Zheng, Xingchen Liu, Zhebin Kuang, Huanxin Yang, Bang Li, Yongge Liu, Lianwen Jin, and Xiang Bai

\section*{\centering Table of Contents}

{\renewcommand{\arraystretch}{1.7}
\begin{tabular}{@{}p{0.97\textwidth}@{}r@{}}
\hyperref[fig:ex-networks]{\textbf{Figure~\ref*{fig:ex-networks}} Detailed network architecture of AlphaOracle framework} \dotfill & \pageref{fig:ex-networks} \\
\hyperref[fig:visualization]{\textbf{Figure~\ref*{fig:visualization}} Case study on oracle bone rubbing} \dotfill & \pageref{fig:visualization} \\
\hyperref[fig:arena_interface]{\textbf{Figure~\ref*{fig:arena_interface}} Interface of the Turing-test-inspired human arena for blind evaluation} \dotfill & \pageref{fig:arena_interface} \\
\hyperref[fig:arena_results]{\textbf{Figure~\ref*{fig:arena_results}} Results of the human arena blind evaluation} \dotfill & \pageref{fig:arena_results} \\
\hyperref[fig:Duizhen]{\textbf{Figure~\ref*{fig:Duizhen}} Visualization of the reading-order parsing result for ``\CJKtrad{屯南} 307''} \dotfill & \pageref{fig:Duizhen} \\
\hyperref[fig:ui-demo]{\textbf{Figure~\ref*{fig:ui-demo}} Interface demonstration of four major functions in AlphaOracle} \dotfill & \pageref{fig:ui-demo} \\
\hyperref[fig:casestudy1]{\textbf{Figure~\ref*{fig:casestudy1}} Case study on OBS} \dotfill & \pageref{fig:casestudy1} \\
\hyperref[fig:casestudy2]{\textbf{Figure~\ref*{fig:casestudy2}} Case study on more OBS cases} \dotfill & \pageref{fig:casestudy2} \\
\hyperref[tab:evaluation_metrics]{\textbf{Table~\ref*{tab:evaluation_metrics}} Detailed Evaluation Metrics and Questionnaire Results} \dotfill & \pageref{tab:evaluation_metrics} \\
\hyperref[tab:interpretation]{\textbf{Table~\ref*{tab:interpretation}} Performance of different language-model backbones on the interpretation task} \dotfill & \pageref{tab:interpretation} \\
\hyperref[tab:alphaoracle-datasets]{\textbf{Table~\ref*{tab:alphaoracle-datasets}} AlphaOracle datasets: processing method, sources and sizes} \dotfill & \pageref{tab:alphaoracle-datasets} \\
\hyperref[sec:supp_method]{\textbf{Appendix A. Supplemental methods}} \dotfill & \pageref{sec:supp_method} \\
\hyperref[sec:data]{\textbf{Appendix B. Data preparation and processing}} \dotfill & \pageref{sec:data} \\
\hyperref[sec:visual_alphaoracle]{\textbf{Appendix C. Visualization of AlphaOracle}} \dotfill & \pageref{sec:visual_alphaoracle} \\
\hyperref[Assessment]{\textbf{Appendix D. Assessment methods}} \dotfill & \pageref{Assessment} \\
\hyperref[sec:detail_expert]{\textbf{Appendix E. Detailed expert profile and evaluation metrics}} \dotfill & \pageref{sec:detail_expert} \\
\hyperref[sec:training_detail]{\textbf{Appendix F. Training details}} \dotfill & \pageref{sec:training_detail} \\
\hyperref[sec:reading_order_parasing]{\textbf{Appendix G. Reading-order parsing for ``Duizhen''}} \dotfill & \pageref{sec:reading_order_parasing} \\
\hyperref[sec:user_interface]{\textbf{Appendix H. User interface}} \dotfill & \pageref{sec:user_interface} \\
\hyperref[sec:limitation]{\textbf{Appendix I. Limitation}} \dotfill & \pageref{sec:limitation} \\
\end{tabular}
}

\section{Supplemental methods}\label{sec:supp_method}
\FloatBarrier
\begin{figure}[!htbp]
\centering
\includegraphics[width=0.98\linewidth]{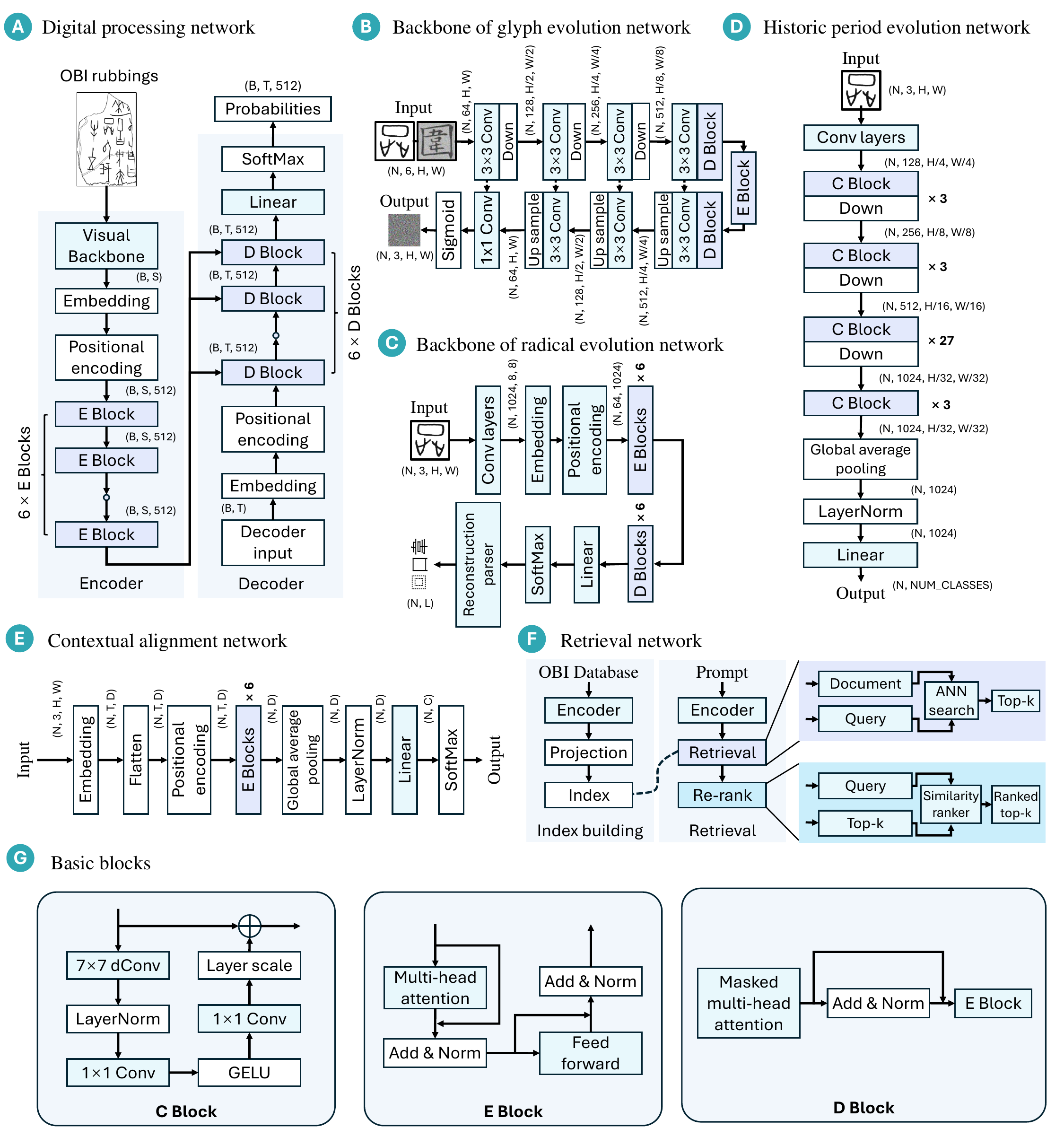}
\caption{\textbf{Detailed network architecture of AlphaOracle framework} (A) Rubbing parsing model for oracle bone rubbing digitization and character detection, featuring visual backbone, embedding layers, and positional encoding. (B) Glyph evolution expert model employing diffusion-based generation with a U-Net architecture for character form evolution modeling. (C) Radical evolution expert model incorporating radical decomposition and reconstruction networks for component-based analysis. (D) Historical period evolution expert Model with convolutional layers and global pooling for cross-period character variant recognition. (E) Contextual alignment metwork employing transformer blocks for distributional semantic analysis of oracle bone script. (F) Retrieval network for philological grounding, integrating document indexing, query processing, and similarity ranking for evidence retrieval from comprehensive scholarly databases. (G) Basic building blocks (C Block, E Block, D Block) showing the fundamental computational units including convolutional layers, attention mechanisms, and normalization components used throughout the framework.}
\label{fig:ex-networks}
\end{figure}

\paragraph{\textbf{\textit{Rubbing parsing model.}}} The \textit{Rubbing parsing model} (Figure~\ref{fig:ex-networks}A) transforms raw oracle bone rubbings into structured character sequences through three coordinated stages: character detection, classification, and sentence reconstruction. This preprocessing is essential because oracle bone inscriptions present unique challenges including irregular spatial layouts, varying character sizes, and the coexistence of deciphered and undeciphered glyphs within the same inscription.

To prioritize system stability, reproducibility, and reliable generalization, we deliberately selected mainstream, battle-tested frameworks for our visual processing pipeline. Character detection employs a DETR-based architecture,\cite{carion2020end} renowned for effectively capturing global context, to accurately localize individual character instances within the irregular and highly complex spatial layouts of the rubbings, producing bounding boxes $b_i \in \mathbb{R}^{4}$ for $i=1,\dots,n$ detected characters. A ConvNeXt-based classifier,\cite{liu2022convnet} extensively validated for its exceptional robustness in complex visual recognition, then processes cropped character regions to reliably assign preliminary categories $s_i$, distinguishing between known characters (which receive specific class labels) and undeciphered instances (which are assigned to a unified placeholder class for downstream analysis by the \textit{Morphological analysis model}).

The critical challenge of sentence reconstruction requires recovering the original reading order from spatially distributed characters. Oracle bone inscriptions follow conventional formulaic patterns but exhibit considerable spatial variation due to the irregular shapes of turtle shells and ox scapulae. We model this as a graph-based ordering problem, where character instances form nodes $V=\{v_1, \cdots, v_n\}$ in a complete graph $G(V,E)$, and potential adjacency relationships are represented as edges $E$. Because the graph is fully connected, adjacency is not restricted by a predefined linear writing direction; instead, the model learns reading transitions for irregular and partially symmetric layouts, including typical ``Duizhen'' cases, directly from annotated spatial patterns. See Supplemental Information for a discussion of ``Duizhen'' cases and a representative example.

A graph neural network processes node features combining category embeddings and positional encodings to predict both character roles (sentence start, middle, or end) and pairwise adjacency relationships. Node features integrate semantic information $c_i = \text{ClassEmbedding}(s_i)$ with spatial information $p_i = \text{PE}(b_i)$, where positional encodings use sine-cosine functions of bounding box coordinates. The network optimizes a combined loss function:
\begin{equation}
L_{\mathrm{sort}} = L_{\mathrm{ce}}(\widetilde{V},V) + 2\,L_{\mathrm{ce}}(\widetilde{E},E),
\end{equation}
where $L_{\text{ce}}$ denotes the standard cross-entropy loss,
$\tilde{V}$ and $\tilde{E}$ represent the predicted node-role distributions
and predicted adjacency edges, respectively,
and $V$ and $E$ denote their corresponding ground-truth labels.
The edge loss receives a higher weight to emphasize correct adjacency prediction. At inference, the system identifies the highest-confidence sentence start and iteratively follows the most probable adjacency edges to reconstruct complete reading sequences.

\paragraph{\textbf{\textit{Morphological analysis model.}}} Paleographers have long emphasized that analyzing the \emph{form} of oracle bone characters is central to their decipherment. This analysis typically proceeds along two complementary axes: (i) \emph{diachronic}, tracing the evolution of graphic elements across historical stages to reveal structural transformations, and (ii) \emph{synchronic}, decomposing characters into constituent radicals whose combinations suggest semantic or phonetic functions.

The computational challenge is to map oracle bone character instances $\mathcal{C}$ to candidate modern equivalents in $\mathbb{M}$, the space of modern Chinese characters. This mapping must account for both evolutionary trajectories across temporal periods and structural decomposition into meaningful components. Oracle bone characters present unique difficulties: extreme graphic variation due to different carvers and media, incomplete or damaged forms, and the absence of phonetic annotations that might guide interpretation.

To address these challenges, the \textit{morphological analysis model} employs an ensemble of three specialized networks, each capturing distinct aspects of character analysis. The \textit{glyph evolution network} (Figure~\ref{fig:ex-networks}B) uses diffusion-based generation, which has been widely proven for its exceptional capability in high-fidelity and highly constrained image generation,\cite{dhariwal2021diffusion} to simulate the historical transformation process from ancient to modern forms, learning patterns of stylistic change across script periods. Due to the inherent randomness of images generated by diffusion models, we set multiple random seeds to generate several Chinese character images. We then use a Chinese character recognition engine to classify these generated images and average the resulting probabilities. The \textit{radical evolution network} (Figure~\ref{fig:ex-networks}C) decomposes characters into constituent elements and reconstructs modern equivalents from these components, enabling inference even when complete characters are unattested. 
The \textit{historic period evolution network} (Figure~\ref{fig:ex-networks}D) incorporates character variants from multiple historical stages, including oracle bone script, Bronze inscriptions, Warring States scripts, Seal script, and Clerical script, into the learning process, and predicts their corresponding modern forms.
Each Expert Model generates probability distributions over candidate modern characters, reflecting different analytical perspectives on the same input glyph. The ensemble integration follows a confidence-weighted combination:
\begin{equation}
p_{\text{ensemble}}(m|\mathcal{C}) = \text{softmax}\left(\sum_{k=1}^{3} w_k \log p_k(m|\mathcal{C})\right),
\end{equation}
where $p_k(m|\mathcal{C})$ represents the probability assigned to modern character $m \in \mathbb{M}$ by the $k$-th Expert Model, and $w_k$ denotes learned model weights. This multi-expert approach emulates the diverse reasoning strategies employed in human paleography, where scholars routinely combine structural, historical, and compositional evidence to build interpretive arguments.

The system operates in two stages: first determining whether detected character instances are already deciphered (by checking consensus across expert models), then generating ranked candidate interpretations for undeciphered instances. Rather than forcing single predictions, the ensemble outputs multiple plausible readings with confidence scores, explicitly acknowledging the inherent uncertainty in morphological analysis and providing robust inputs for downstream contextual validation.

\paragraph{\textbf{\textit{Contextual alignment model.}}} The \textit{contextual alignment model} exploits these contexts through large-scale retrieval of related inscriptions, followed by neural contextual modeling (Figure~\ref{fig:ex-networks}E). Before analysis, the system retrieves all available occurrences of target characters from comprehensive digitized corpora, including \emph{Jiaguwen Heji}, \emph{Jiaguwen Moben Daxi}, and \emph{Jiaguwen Jiaoshi Zongji}, ensuring that interpretations are evaluated against the fullest possible inscriptional evidence.

The contextual modeling component employs a BERT-based architecture with transformer blocks,\cite{devlin2019bert,tranformer} chosen for its proven ability to capture deep bidirectional representations. The masked language modeling training strategy naturally parallels oracle character restoration scenarios. Given oracle sentences with unknown characters marked as \texttt{<MASK>} tokens, the model predicts probability distributions over the modern Chinese vocabulary $\mathbf{V}$ containing 23,292 characters, given the input sequence X representing the entire oracle-bone sentence:
\begin{equation}
P(\hat{y}=v \mid \mathbf{X}), \quad v \in \mathbf{V},
\end{equation}

Training follows a masked language modeling paradigm where a fraction of tokens in deciphered oracle sentences are masked, and the model learns to predict them from surrounding context. The training loss optimizes:
\begin{equation}
L = -\sum_{i=1}^{n} w_i \log P(\hat{y}_i = c_i \mid \mathbf{X}),
\end{equation}
where $w_i$ indicates whether token $c_i$ is masked for prediction.

At inference, the model produces two complementary outputs for undeciphered characters: contextual rescoring of morphological candidates based on semantic plausibility, and independent semantic predictions from the full vocabulary:
\begin{equation}
\left\{
\begin{alignedat}{1}
& r_{\mathrm{score}}    = \mathrm{TopK}\!\bigl(\mathbf{P}(m_{1}, \dots, m_{N}), n_{\mathrm{score}}\bigr) \\
& r_{\mathrm{semantic}} = \mathrm{TopK}\!\bigl(\mathbf{P}(\mathbf{V}), n_{\mathrm{semantic}}\bigr),
\end{alignedat}
\right.
\end{equation}
where $\{m_1,\dots,m_N\}$ morphological candidates and $\mathbf{P}$ denotes the model's probability distribution.

The system integrates an interpretation component that generates modern-language explanations through a two-stage training strategy. Supervised fine-tuning (SFT) teaches structured reasoning with explicit separation between analytical steps (\texttt{<think>}--\texttt{</think>}) and final interpretations (\texttt{<answer>}--\texttt{</answer>}).  Training data includes multiple reasoning paths for each inscription, enabling articulation of alternative analytical chains. Reinforcement learning with feedback-derived rewards further improves generalization by optimizing a KL-regularized PPO-style objective that balances lexical accuracy, measured by BLEU-4 and ROUGE-L, with format consistency.\cite{papineni2002bleu}

This dual training approach enables the system to aggregate interpretations across inscriptions containing identical characters, identifying recurring semantic patterns and variations across contexts. The transparent reasoning chains facilitate expert evaluation while significantly reducing manual corpus search effort. Through this integrated approach, the contextual alignment model ensures that character interpretations are grounded not only in visual similarity but also in actual usage patterns observed across the oracle bone corpus. This contextual validation significantly reduces the interpretive space and provides essential evidence for downstream philological analysis.

\paragraph{\textbf{\textit{Philological grounding model.}}} Given a candidate interpretation for a target glyph (from morphological analysis and contextual alignment), the module
(i) \emph{expands the query} using known variant forms, graphic components, and near-synonyms;
(ii) \emph{retrieves passages} from the indexed corpus via our semantic vector retrieval model and glyph image retrieval model
(iii) \emph{normalizes and deduplicates} hits across editions, aligning to canonical references; and
(iv) \emph{filters by period and genre} to privilege sources with higher evidential value (e.g., pre-Qin prose over much later commentaries for core semantics).
Each passage is stored together with bibliographic metadata, time period, a genre tag, a brief rationale indicating its relevance, as well as the structured text produced by PDF parsing.

Semantic vector retrieval is instantiated with Qwen3-Embedding,\cite{zhang2025qwen3} with passage vectors stored in a Qdrant collection configured with an HNSW index and cosine similarity. Glyph image retrieval employs DINO v2,\cite{oquab2023dinov2} extracting multi-scale visual descriptors from binarized glyph crops and conducting nearest-neighbor search with the L2 metric under per-script normalization.

Retrieved passages $T$ are aggregated into a literature-support score: 

\begin{equation}
S_{\operatorname{phil}}(T \mid q) = \sum_{c \in T} \alpha_{c,q} \operatorname{sim}(c, q) \operatorname{auth}(T),
\end{equation}
where $c$ is the candidate under evaluation, $q$ denotes the user query (textual $q^{\text{text}}$ for the semantic model or visual $q^{\text{img}}$ for the glyph model), $\alpha_{c,q}$ utilizes Qwen3-Reranker for a more fine-grained calculation of relevance and reliability (in image retrieval, $\alpha_{c,q}$ is always set to 1), $\operatorname{auth}(T)$ denotes the authority of the passage $T$, and $\operatorname{sim}(c, q)$ uses stored passage vectors to rapidly calculate the similarity between $c$ and $q$. To ensure precise and consistent semantics, we define $\operatorname{sim}$ by model family:

\begin{equation}
\operatorname{sim}(c,q)
\;=\;
\begin{cases}
\operatorname{sim}_{\text{semantic}}\!\bigl(c, q^{\text{text}}\bigr),
& \text{(semantic retrieval)}\\[4pt]
\operatorname{sim}_{\text{glyph}}\!\bigl(c, q^{\text{img}}\bigr),
& \text{(glyph retrieval)},
\end{cases}
\end{equation}
where \emph{semantic vector retrieval model} measures the similarity between the selected span of passage $T$ and the textual query $q^{\text{text}}$, using context-aware embeddings with synonym and variant-form expansion. The \emph{glyph image retrieval model} measures the graphical similarity between the selected glyph image within passage $T$ and the query image $q^{\text{img}}$, based on visual features of the character glyph.

\paragraph*{\textbf{\textit{Comprehensive report.}}}
The \textit{comprehensive report} represents the final stage of the pipeline. It compiles the pre-computed outputs of all preceding modules into a structured decipherment report for each undeciphered oracle-bone character. This component performs no additional scoring or adjudication; rather, it functions as an integrative reporting interface that organizes and elucidates the collective findings.

For each character, the \textit{comprehensive report} receives the complete set of upstream outputs: the initial character sequence and positional information from the rubbing parsing model; candidate modern readings, confidence scores, and glyph-evolution evidence from the \textit{morphological analysis model}; the contextually restored sequence, modern-language interpretation, and inferred semantic meaning from the \textit{contextual alignment model}; and the retrieved parallel inscriptions and corroborating historical texts from the \textit{philological grounding model}.

Using these inputs, the \textit{comprehensive report} presents three layers of synthesis. First, it summarizes glyphic evidence by integrating candidate readings, confidence scores, structural comparisons, and evolutionary patterns. Second, it explains the contextual evidence by describing how the candidate readings align with the restored inscriptional sequence and its inferred semantic meaning. Third, it consolidates broader semantic patterns and external philological validation by incorporating supporting inscriptions and relevant historical sources.

The report concludes with an overall assessment that identifies the most strongly supported interpretation and its associated confidence level, while explicitly acknowledging any remaining uncertainties. Through this process, the structured decipherment report provides epigraphers and historians with a transparent, coherent, and rigorously organized synthesis.

\section{Data preparation and processing}\label{sec:data}

To support large-scale computational analysis, we constructed the most comprehensive unified corpus of oracle bone inscriptions.

Since the majority of oracle bone–related materials are preserved in forms that are difficult to use directly (such as PDFs or image documents), we designed a complete digitization pipeline following a paradigm similar to MonkeyOCR.\cite{li2025monkeyocr} First, we perform coarse-grained layout detection and reading-order analysis based on PP-StructureV3.\cite{cui2025paddleocr} Second, we carry out fine-grained character detection, classifying text lines into three categories: OBS, modern Chinese characters, and other symbols. Finally, we apply separate recognition strategies: OBS characters are stored as images for subsequent training and retrieval; modern Chinese characters are processed with a radical-based recognition model to enhance generalization across nearly 90,000 character classes; and other symbols are handled with a dedicated classifier. Through this pipeline, we converted the annotations of oracle bone rubbings, their modern Chinese translations, as well as \textit{Gulin} and other authoritative modern scholarly studies, from PDF and image formats into machine-readable structured data.

Specifically, for the annotations of rubbings and transcriptions, we collected recognition texts from the digitized \emph{Jiaguwen Moben Daxi} and \emph{Jiaguwen Jiaoshi Zongji}. In addition, we obtained image–ID pairs for transcriptions from \emph{Jiaguwen Moben Daxi} using OpenCV, and collected image–ID pairs for rubbings from \emph{Jiaguwen Heji} through web crawling, and further incorporated publicly available data from the \emph{OBIMD} dataset.\cite{li2024oracle}
For the modern language translation component, we assembled a curated corpus of inscription–translation pairs drawn from authoritative sources, including scholarly translation volumes, expert-annotated corpora, and classical Chinese interpretive texts.
These pairs were further expanded by incorporating reasoning steps, enabling the development of models that do not merely produce translations but also reconstruct the interpretive process. For modern scholarly research, we digitized \emph{Guwenzi Gulin} and \emph{Jiaguwenzi Gulin}, which contain expert interpretations of ancient scripts, along with contemporary research papers. For ancient transmitted texts, we additionally incorporated and processed the open-source datasets \emph{Classical-Modern} and \emph{Shuowen Jiezi Dataset}. For OBS glyphs, we used our previously released datasets \emph{HUST-OBC} and \emph{EVOBC},\cite{wang2024open,guan2024open} as well as the publicly available dataset \emph{OBIMD}. All dataset names, data volumes, and processing methods are provided in Table \ref{tab:alphaoracle-datasets}.

\section{Visualization of AlphaOracle}\label{sec:visual_alphaoracle}
\begin{figure}[t!]
    \centering
    \includegraphics[width=0.87\linewidth]{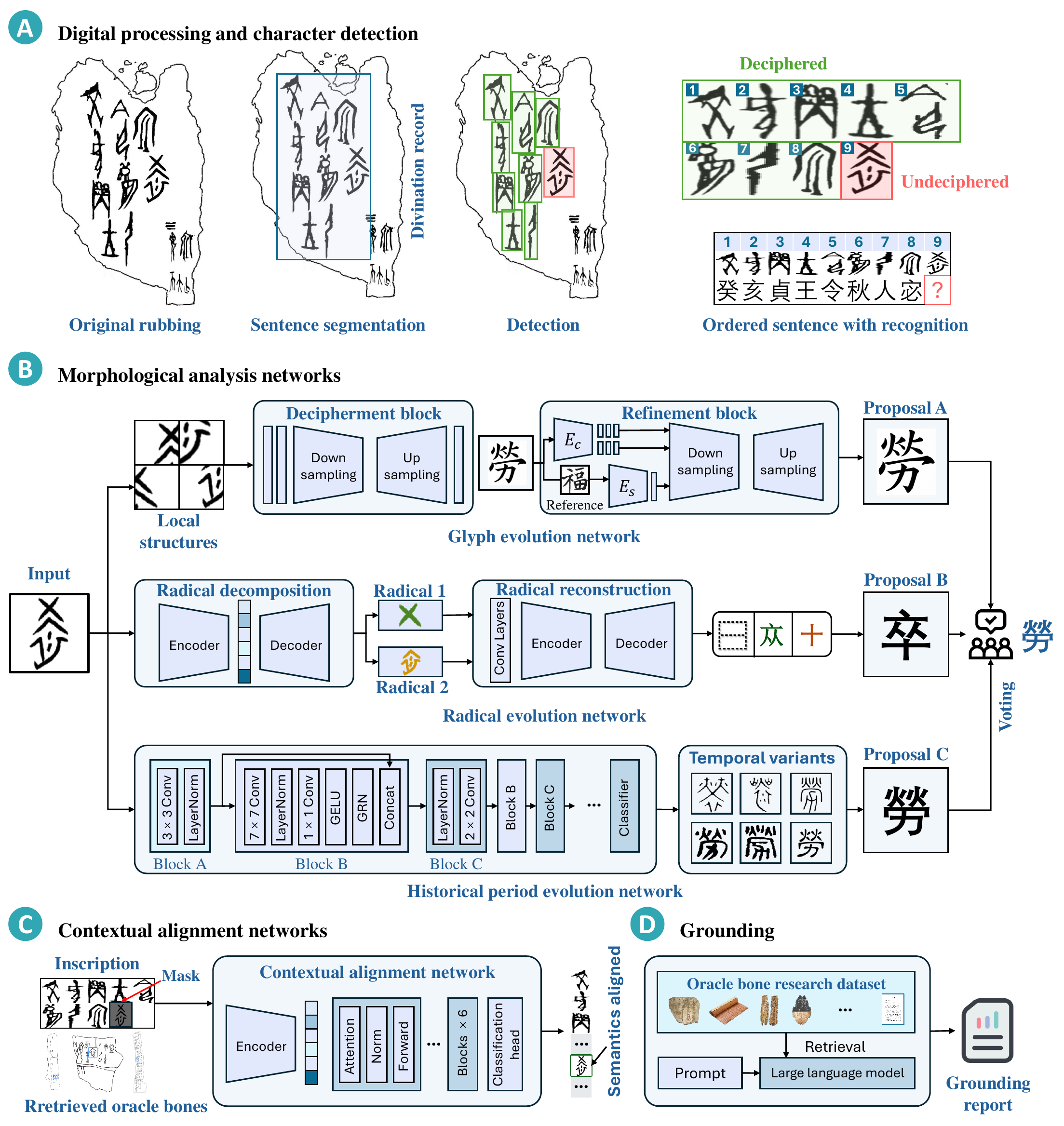}
    \caption{
    \textbf{Case study on oracle bone rubbing ``\texttt{\CJKtrad{屯}} 307''}
(A) Digital processing and character detection. 
(B) Morphological analysis with expert models. 
(C) Contextual alignment. 
(D) Decipherment report generation for the contested character.
}
    \label{fig:visualization}
    \vspace{-0.2cm}
\end{figure}

To illustrate the complete decipherment workflow of AlphaOracle, we present a detailed case study of oracle bone rubbing ``\CJKtrad{屯} 307'' from the \textit{Xiaotun Nandi OBS}. As shown in Figure~\ref{fig:visualization}A, this rubbing preserves three divination records: two are relatively brief, whereas the third presents a more complex sentence and is well suited to illustrating our analysis workflow. The same example further illustrates reading-order recovery under a partially symmetric “Duizhen” layout, as shown in Figure \ref{fig:Duizhen}.

The rubbing parsing model systematically segments the original rubbing into three divination records and already parsing nine characters in the first complex sentence. Eight are recognized as known forms, while position 7 (``\raisebox{-0.4ex}{\includegraphics[height=2.5ex]{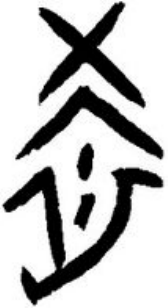}''} in Figure \ref{fig:visualization}B) remains disputed. Subsequently, AlphaOracle will undergo a detailed analysis using three modules: morphological analysis, contextual alignment, and philological grounding.

The morphological analysis stage reveals remarkable convergence across all three expert models (see Figure~\ref{fig:visualization}B). The glyph evolution network detects structures consistent with the semantics of labour, suggesting work by lamplight. The radical evolution network decomposes the graph into recognisable parts and assigns it to “\CJKtrad{卒}”. The historical period evolution network traces cognate forms in bronze inscriptions and later scripts, linking them closely to “\CJKtrad{勞}”. Despite the radical network’s discordant call, the three-model ensemble favours “\CJKtrad{勞}” as the most parsimonious reading, providing a strong preliminary consensus for contextual verification. Early scholarship typically read the graph as “\CJKtrad{卒}”, but recent revisions favour “\CJKtrad{勞}”.

As shown in Figure~\ref{fig:visualization}C, contextual alignment transforms this morphological hypothesis into a corpus-wide investigation. The system retrieves ten oracle bone fragments from the comprehensive digitized collections (specimens 33473, 33535, 34656, 34732, 34781, 34800, 35554, 35560, 35561, and 35562) containing visually identical or highly similar character forms. By extracting the relevant sentences and applying masked semantic alignment, the module tests whether ``\CJKtrad{勞}'' produces coherent interpretations across diverse contexts. These examples almost all point to the same usage ``\CJKtrad{在\raisebox{-0.4ex}{\includegraphics[height=2.5ex]{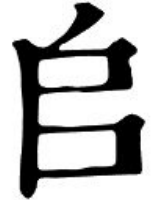}}勞卜}" (\textit{divination was performed in the tall ancestral temple at [Lao]}). The target inscription ``\CJKtrad{屯}307'' itself reconstructs as ``\CJKtrad{令秋人宓勞}" (\textit{Instruct the Qiu clan to subdue the area of Lao}). The term is consistently used as a toponym or ethnonym across all retrieved contexts. In conjunction with the morphological analysis, this suggests a social reality in which the local tribe required pacification or relief due to popular hardship.

The philological grounding module (see Figure~\ref{fig:visualization}D) extends this analysis by consulting authoritative scholarly resources, retrieving the top sixteen relevant studies from the comprehensive literature database. These include \textit{Guwenzi Gulin, Volume 10, Page 429} Qiu Xigui's \textit{Explanation of Bi}, and Pan Yunzhong's \textit{The Formation and Development of Basic Chinese Vocabulary}. The retrieved literature suggests that “\CJKtrad{勞}” in oracle bone inscriptions primarily functioned as a toponym. Over the course of historical evolution, its meaning gradually developed into that of diligence and labor. Furthermore, Qiu Xigui argues that ``\raisebox{-0.4ex}{\includegraphics[height=2.5ex]{48th-AlphaOra-Report-1893-InlineGlyph-Lao.pdf}}'' in oracle bone inscriptions cannot be simply identified as ``\CJKtrad{卒}''.

This systematic evidence integration demonstrates how AlphaOracle transforms interpretive challenges into structured analytical workflows. Rather than relying on isolated visual similarity or expert intuition, the framework assembles converging evidence from morphological plausibility, distributional consistency, and philological continuity. The resulting interpretation of \CJKtrad{屯}307 is that it records a divination concerning the Shang king’s order for the Qiu people to pacify a place called Lao, illustrating how computational analysis can illuminate both linguistic structure and historical content. By maintaining explicit documentation of analytical steps and uncertainty levels, the system provides a transparent foundation for scholarly evaluation while advancing understanding of Shang-period strategic concerns and religious practices. Representative user-interface views for rubbing processing, morphological analysis, contextual alignment, and philological grounding are provided in Figure \ref{fig:ui-demo}.

\begin{figure}[!htbp]
    \centering
    \includegraphics[width=0.87\linewidth]{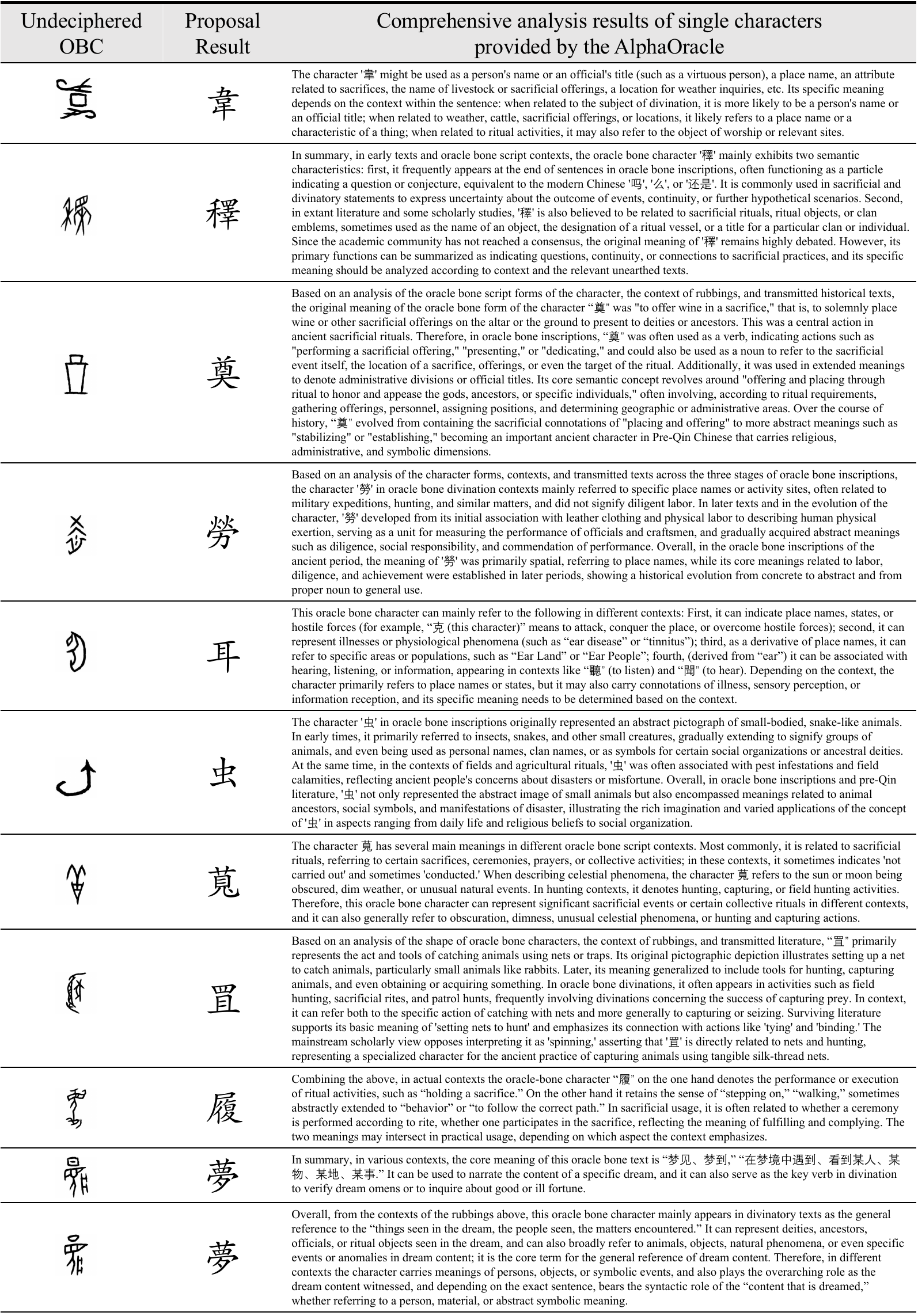}
    \caption{
    \textbf{Case study on OBS}}
    \label{fig:casestudy1}
    \vspace{-0.2cm}
\end{figure}

\begin{figure}[!htbp]
    \centering
    \includegraphics[width=0.87\linewidth]{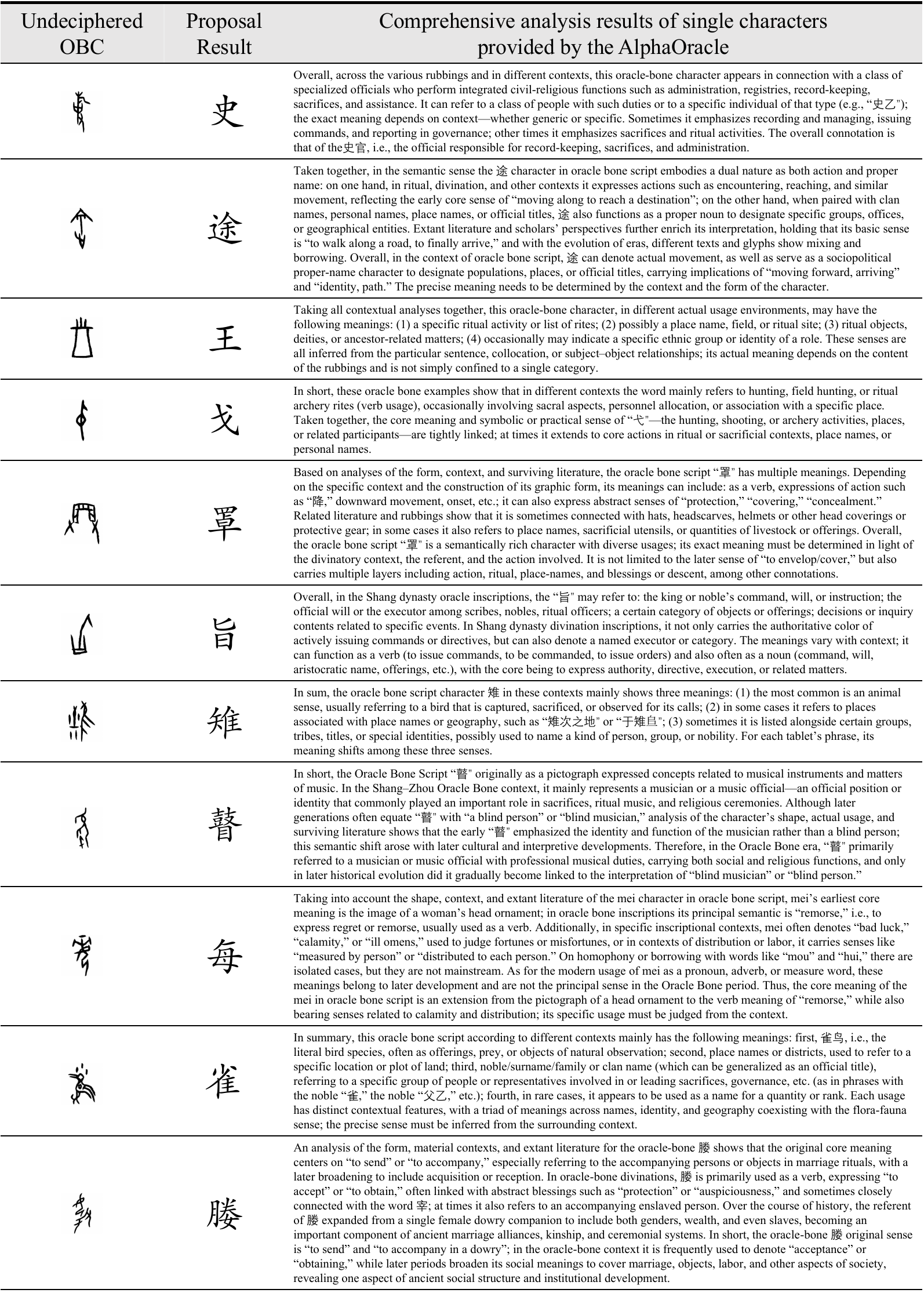}
    \caption{
    \textbf{Case study on more OBS cases}}
    \label{fig:casestudy2}
    \vspace{-0.2cm}
\end{figure}

Furthermore, AlphaOracle disambiguates 22 additional contested oracle bone variants, offering fresh insights into the writing practices and interpretative traditions of the Shang dynasty (see Figure~\ref{fig:casestudy1} and Figure~\ref{fig:casestudy2}).

\section{Assessment methods} \label{Assessment}

This study aims to simulate an authentic oracle bone script decipherment workflow. The assessment process consists of three complementary setups. 

\paragraph{\textbf{\textit{Morphological analysis.}}} To evaluate the model’s ability to recognize and generalize to previously unseen characters, we partitioned the dataset into a training set, a validation set, and a test set based on OBS character categories (with a ratio of 9:0.5:0.5). First, the model was trained on the training set to capture morphological features of known characters. Second, ensemble selection was guided by performance on the validation set, which contained character classes not present during training. Finally, the model’s core decipherment capability was verified on the test set, also consisting of unseen characters. Specifically, the dataset was split as follows: \emph{TRAIN:} 1577 categories, 148,730 samples; \emph{VAL:} 87 categories, 6,748 samples; \emph{TEST:} 88 categories, 8,032 samples.

\paragraph{\textbf{\textit{Contextual alignment.}}} Constrained by the vocabulary size of the language model, we divided the dataset into a training set and a test set according to sample counts. The model was trained to perform contextual alignment on the training set, and its accuracy was then measured on the test set. This allowed systematic assessment of the system’s ability to leverage inscriptional context in supporting character interpretation.
To evaluate the capability of the interpretation model, the Jiaguwen Jingcui Shiyi Dataset was split into a training set and a test set, with 200 samples for testing and 1,300 samples for training. First, the model was fine-tuned on 1,200 samples using SFT, enabling it to initially acquire the ability to perform interpretation based on a chain-of-thought. Next, 100 samples were used with reverse GRPO to enhance the model’s stability on unseen data. Finally, the model’s performance on the test set was measured using BLEU, ROUGE, and METEOR metrics.

\paragraph{\textbf{\textit{Expert evaluation.}}} Beyond automated benchmarks, we conducted an expert study to assess the system’s practical value under real scholarly review. A panel of specialists (86) in oracle bone studies participated anonymously to protect privacy. Experts engaged with the system across multiple tasks, including analyzing rubbings containing undeciphered characters, evaluating candidate glyph interpretations, and examining retrieval results from the contextual and philological modules. Feedback was collected both on the plausibility of proposed readings and on the usability of evidence bundles generated by the system. Experts were also asked to assess the extent to which the system could reduce the time required for comparable analyses, in order to evaluate its potential to improve efficiency in scholarly workflows. This evaluation complements quantitative metrics by aligning assessment with the expectations and practices of professional paleographers.

\paragraph{\textit{Human evaluation for interpretation quality.}} To further strengthen the objectivity of interpretation evaluation, we introduce an additional blind human assessment protocol, termed the \textbf{human arena}. This evaluation specifically targets the \emph{accuracy} of generated interpretations while implicitly penalizing \emph{over-interpretation}.

\begin{figure}[ht!]
    \centering
    \includegraphics[width=0.8\linewidth]{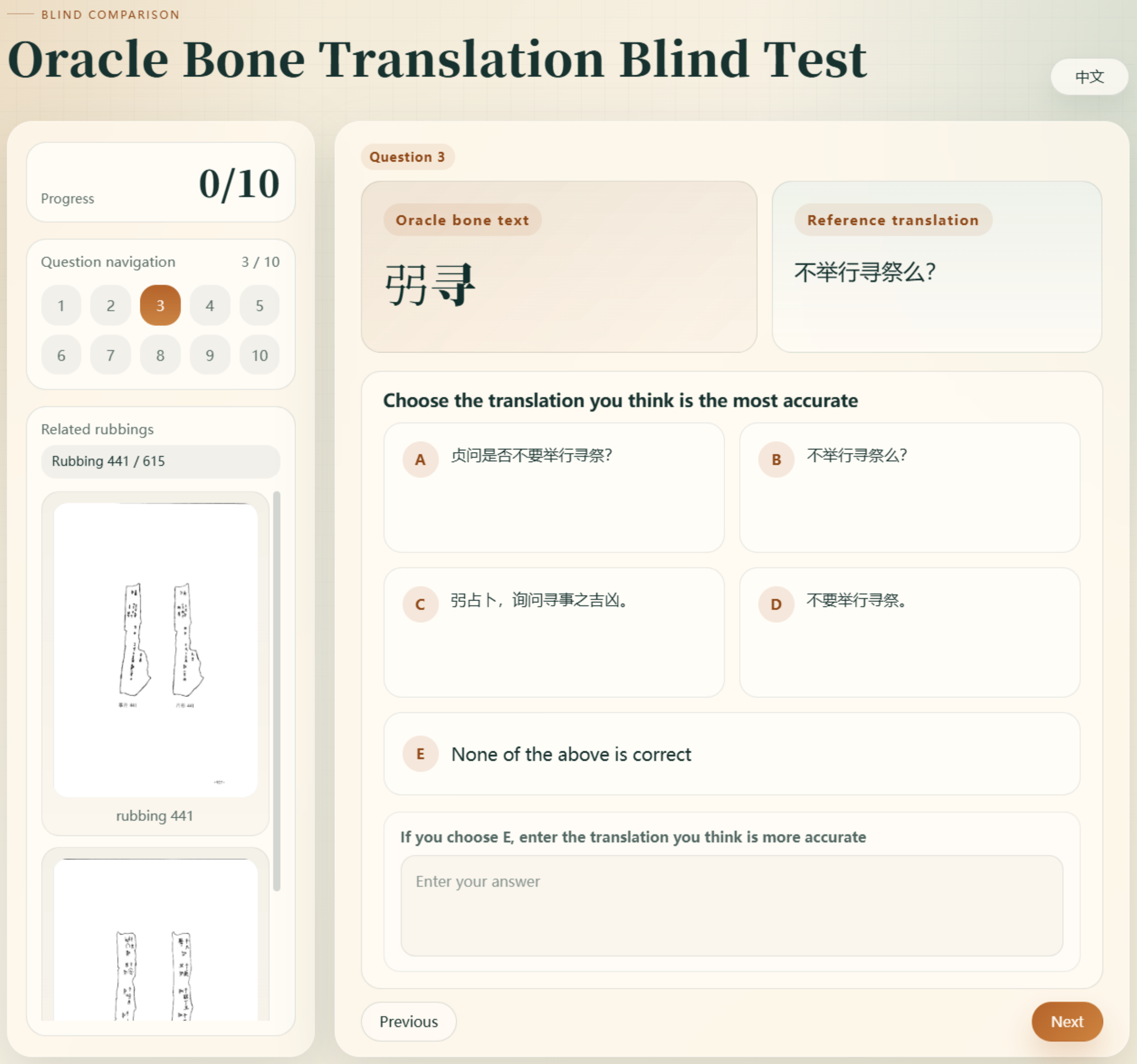}
    \caption{\textbf{Interface of the Turing-test-inspired human arena for blind evaluation} Experts are presented with the original oracle bone rubbing, a reference transcription, and four anonymized candidate interpretations in randomized order. An optional free-text field allows experts to provide their own readings.}
    \label{fig:arena_interface}
\end{figure}

The evaluation follows a Turing-test-inspired design. As shown in Figure~\ref{fig:arena_interface}, each expert is presented with the original oracle bone rubbing image and its corresponding reference transcription. Four candidate interpretations, generated respectively by GPT-5, Gemini 2.5 Pro, DeepSeek, and our method, are displayed in a randomly shuffled order with anonymous attribution. Experts are required to select the interpretation that best matches the reference semantics. No external materials or digital tools are permitted during evaluation, ensuring a strictly blind setting free from source bias.

To ensure the rigor and authority of the human evaluation, we invited \textbf{21 senior scholars} specializing in oracle bone script research to participate. Each expert independently evaluated 10 randomly sampled interpretation instances, yielding a total of 210 judgments.

\begin{figure}[ht!]
    \centering
    \includegraphics[width=0.8\linewidth]{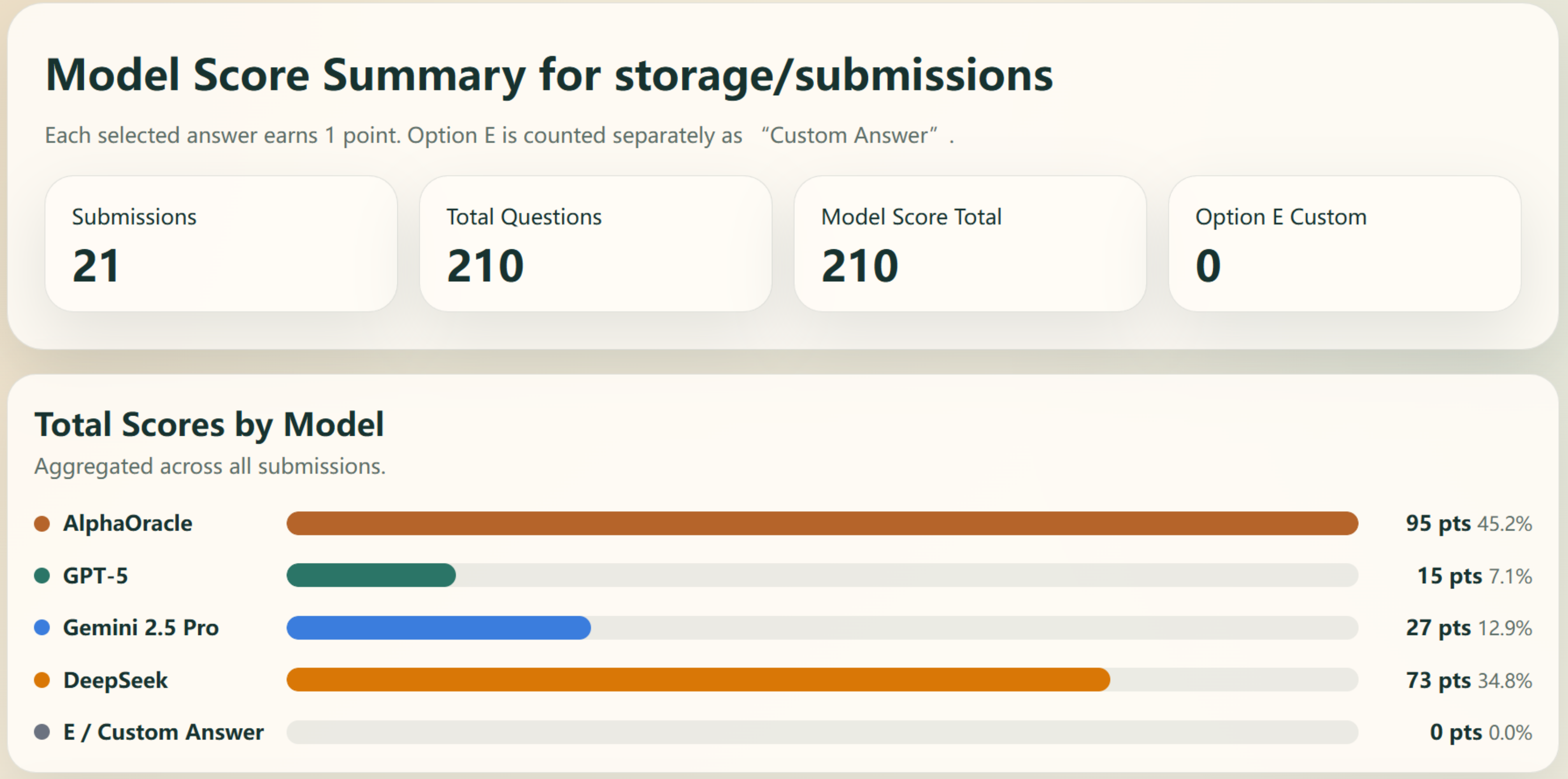}
    \caption{\textbf{Results of the human arena blind evaluation} Evaluation by 21 senior oracle bone script scholars (210 total judgments). Our method achieves the highest expert preference rate of 45.2\%.}
    \label{fig:arena_results}
\end{figure}

The results are presented in Figure~\ref{fig:arena_results}. Our method achieved the highest preference rate of \textbf{45.2\%}, followed by GPT-5 (7.1\%), Gemini 2.5 Pro (12.9\%), and DeepSeek (34.8\%). This demonstrates that our system's interpretations are most consistently aligned with expert-judged quality standards among all compared methods.

It is also worth reporting the qualitative feedback provided by the participating scholars, which offers valuable insights into the current limitations of all systems:

\begin{itemize}
    \item Several experts noted that for a small number of inscriptions, none of the four model outputs were satisfactory. One expert specifically recommended consulting pp.\ 160--166 of \emph{Collected Works of Qiu Xigui: Oracle Bone Inscriptions Volume} for more nuanced readings of certain contested characters.
    \item Some experts observed that certain candidate interpretations exhibited no substantive differences from one another.
    \item Multiple scholars pointed out that a considerable number of oracle bone inscriptions still lack scholarly consensus on their readings, reflecting the inherent ambiguity of the source material rather than a deficiency of any particular model.
    \item Experts suggested adding an option for evaluators to input their own interpretation when none of the candidates is deemed adequate. We have incorporated this as an optional free-text field in the revised evaluation interface.
    \item Notably, some experts observed that certain reference texts from \emph{Erta Essence Interpretations}  themselves contain errors. When the reference transcription is incorrect, the ``translation'' inevitably follows suit (regardless of the model used). This reveals the compounding effect of upstream annotation quality on downstream evaluation.
\end{itemize}

These observations collectively affirm that our system is \textbf{not intended to replace domain experts}, but rather to serve as a powerful assistive tool. From a technical standpoint, the Human Arena results provide strong evidence that our method achieves meaningful progress in oracle bone script interpretation, producing outputs that domain specialists recognize as the highest quality among current automated approaches.

\section{Detailed expert profile and evaluation metrics}\label{sec:detail_expert}
During the testing phase of AlphaOracle, 86 domain experts and doctoral researchers participated in evaluating the system. To encourage candid and unbiased feedback, the survey was conducted under strict anonymity and data privacy commitments. Given the highly specialized and concentrated nature of the oracle bone research community, disclosing granular demographic data (such as specific institutional affiliations or exact years of experience) could inadvertently compromise participant anonymity. However, to ensure a balanced evaluation, we recorded the broad research domains of the 86 evaluators. Furthermore, the evaluation metrics assessed the system's overall helpfulness, efficiency improvement, and performance across four core functionalities using a 5-point Likert scale, where 1 is the lowest and 5 is the highest.

The detailed professional domain distribution of the experts and the aggregated questionnaire results are comprehensively presented in Table \ref{tab:evaluation_metrics}.

\begin{table}[H]
\centering
\renewcommand{\arraystretch}{1.05}
\setlength{\tabcolsep}{4pt}
\small
\caption{Detailed evaluation metrics and questionnaire results}
\label{tab:evaluation_metrics}
\makebox[\textwidth][c]{%
\begin{tabular}{p{0.38\textwidth} p{0.55\textwidth}}
\toprule
\textbf{Evaluation Metric /} \newline \textbf{Questionnaire Item} & \textbf{Evaluation Results \& Distribution \hspace*{\fill} (N=86)} \\
\midrule
\textbf{Q1:} Field of Work & 
Oracle Bone/Ancient Texts: 36.17\% (31) \newline
Computer Science/AI: 38.30\% (33) \newline
Other Humanities/Social Sciences: 17.02\% (15) \newline
Other STEM: 8.51\% (7) \\
\midrule
\textbf{Q2:} Helpfulness for \newline deciphering Oracle Bones & 
Very helpful: 40.43\% (35) \newline
Quite helpful: 38.30\% (33) \newline
Somewhat helpful: 19.15\% (16) \newline
Unclear/Cannot evaluate: 2.13\% (2) \newline
Little to no help / No help: 0\% (0) \\
\midrule
\textbf{Q3:} Efficiency improvement \newline / Time saved & 
80\%-100\% efficiency gain: 23.40\% (20) \newline
70\% efficiency gain: 44.68\% (38) \newline
50\% efficiency gain: 17.02\% (15) \newline
30\% efficiency gain: 10.64\% (9) \newline
10\% efficiency gain: 2.13\% (2) \newline
0\% efficiency gain: 2.13\% (2) \\
\midrule
\textbf{Q4:} Rating for ``Rubbing \newline Copy Analysis'' (1-5) & 
\textbf{Avg: 4.20} ~--~ 5: 46.81\% (40), ~ 4: 38.30\% (33), \newline 3: 8.51\% (7), ~ 2: 2.13\% (2), ~ 1: 4.26\% (4) \\
\midrule
\textbf{Q5:} Rating for ``Single \newline Character Analysis'' (1-5) & 
\textbf{Avg: 4.44} ~--~ 5: 57.45\% (49), ~ 4: 31.91\% (28), \newline 3: 8.51\% (7), ~ 2: 2.13\% (2), ~ 1: 0\% (0) \\
\midrule
\textbf{Q6:} Rating for ``Single \newline Character Retrieval'' (1-5) & 
\textbf{Avg: 4.37} ~--~ 5: 57.45\% (49), ~ 4: 25.53\% (22), \newline 3: 14.89\% (13), ~ 2: 2.13\% (2), ~ 1: 0\% (0) \\
\midrule
\textbf{Q7:} Rating for ``Related \newline Literature Query'' (1-5) & 
\textbf{Avg: 4.01} ~--~ 5: 42.55\% (37), ~ 4: 31.91\% (27), \newline 3: 14.89\% (13), ~ 2: 4.26\% (4), ~ 1: 6.38\% (5) \\
\bottomrule
\end{tabular}
}
\end{table}

\section{Training details}\label{sec:training_detail}
To ensure the reproducibility of AlphaOracle, we provide the implementation details of each component as follows.

\textbf{\textit{Rubbing parsing model}} consists of rubbing parsing model and glyph analysis model.
In rubbing parsing model, the detection and recognition model was trained on OBIMD dataset and Oracle Character Detection Dataset to enhance recognition robustness across different rubbing styles. It used ResNet-50 as the backbone, with a decoder consisting of 6 layers and 900 query tokens. The optimizer was AdamW, with an initial learning rate of $2\times 10^{-4}$, a weight decay of $1\times 10^{-4}$, and a learning rate factor of 0.1. Gradient clipping was applied with $L_2 \leq 0.1$. The learning rate decayed by a factor of 0.1 at the 30th epoch following a Multi-Step schedule, and remained unchanged during other stages. Training was conducted for a total of 100 epochs. Input images of size $1024\times 1024$ were processed using Large-Scale Jitter (LSJ) augmentation with random scaling, cropping, flipping, and fixed white padding. Moreover,the sorting model was trained on OBIMD dataset and used a fixed learning rate of $1\times 10^{-4}$ during the optimization stage, with a per-GPU batch size of 4, resulting in a total effective batch size of 32. The hidden feature dimension of the main network was set to 128, and the output dimension was also 128. The graph learning module applied regularization terms with $\gamma = 1\times 10^{-4}$ and $\eta = 1.0$, and introduced a 128-dimensional graph representation projection space. The message-passing component adopted a 2-layer GCN structure to fully capture the graph topology information. The model was trained for a total of 100 epochs.

\textbf{\textit{Morphological analysis models}} is trained on HUST-OBC, EVOBC and OBIMD datasets to ensure comprehensive coverage of diverse glyph styles and variations. Among them, OBSD was trained using the Adam optimizer with a weight decay of $10^{-4}$, $\beta_1 = 0.9$, and $\beta_2 = 0.999$. During training, the learning rate was set to $2\times 10^{-5}$, and the batch size was 8. Each batch contained 8 patches of size $64\times 64$, and the model was trained for 300 epochs. The entire training process spanned over 2 weeks. And P$^3$ used ResNet-50 as the foundational backbone network. For the Transformer components, both the encoder and decoder were structured with six layers and featured eight attention heads. Optimization was achieved through the AdamW optimizer, starting with an initial learning rate of $5\times 10^{-4}$, which was linearly decreased to a minimum of $1\times 10^{-6}$. The model was trained with a batch size of 256 for 100 epochs. Meanwhile, EVCS used the AdamW optimizer with an initial learning rate $lr = 1\times 10^{-3}$, a weight decay coefficient of $0.05$, and momentum parameters $\beta_1 = 0.9$ and $\beta_2 = 0.999$. The learning rate underwent a linear warm-up over the first 20 epochs (starting from 0.1\%), and subsequently decayed to a minimum value of $1\times 10^{-5}$ following a cosine annealing schedule. Training was conducted for a total of 300 epochs; each mini-batch consisted of 108 images of size $128\times 128$, with data augmentation strategies including random scaling, cropping, flipping, and illumination perturbations.

\textbf{\textit{Contextual alignment model}} is composed of contextual restoration model and interpretation model.
Contextual restoration model trained on the \textit{OBIMD} and \textit{Jiaguwen Moben Daxi} Dataset. The model adopted the AdamW optimizer with a weight decay coefficient of $0.01$ and an initial learning rate of $3\times 10^{-4}$. A cosine learning rate scheduling strategy was used to gradually decay the learning rate throughout the training process, with a total of 50 epochs. For masked language modeling, 15\% of the tokens in the input sequence were selected for prediction, of which 80\% were replaced with the [MASK] token, 10\% were replaced with random tokens from the vocabulary, and the remaining 10\% were left unchanged. 
Interpretation Model used Jiaguwen Jingcui Shiyi Dataset for training. For the main system, we selected Qwen2.5-7B as the default base model. To assess backbone robustness, we additionally instantiated the same interpretation pipeline with Llama3.1-8B-Instruct and Ministral-8B-Instruct. The training processes were completed based on LLaMA-Factory and open-r1 for SFT and GRPO, respectively. Unless otherwise noted, the same data split, optimization recipe, and decoding settings were used across all three backbones. During training, a batch size of 8 and 10 training epochs were set for SFT, and a batch size of 16 and 40 training epochs were set for GRPO. The inference temperature was set to 0.9, allowing parallel generation of 16 candidate answers. A fixed random seed of 42 was used for reproducibility. The experimental results for the three models are shown in Figure \ref{tab:interpretation}.

\begin{table}[H]
\centering
\renewcommand{\arraystretch}{1.1}
\setlength{\tabcolsep}{4pt}
\small
\caption{Performance of different language-model backbones on the interpretation task}
\label{tab:interpretation}
\begin{tabular}{lccc}
\toprule
\textbf{Model} & \textbf{BLEU} & \textbf{ROUGE} & \textbf{METEOR} \\
\midrule
Qwen2.5-7B & 0.491 & 0.586 & 0.659 \\
Llama3.1-8B-Instruct & 0.441 & 0.547 & 0.614 \\
Ministral-8B-Instruct & 0.418 & 0.520 & 0.596 \\
\bottomrule
\end{tabular}
\end{table}

\textbf{\textit{Philological grounding model}} is trained using a curated corpus comprising \textit{OBIMD}, Classical-Modern, \textit{Shuowen Jiezi} dataset, \textit{Jiaguwen Moben Daxi} dataset, \textit{Guwenzi Gulin} dataset, \textit{Jiaguwenzi Gulin} dataset, \textit{Jiaguwen Jiaoshi Zongji}, and \textit{Jiaguwen Heji} dataset. The model adopted the AdamW optimizer ($\beta_{1} = 0.9$, $\beta_{2} = 0.95$) with an initial learning rate of $1\times 10^{-4}$ and a weight decay coefficient of $0.04$. The learning rate followed a cosine annealing schedule, decaying from $1\times 10^{-4}$ to $1\times 10^{-6}$, with a warm-up period of 10 epochs; the weight decay was updated by a synchronized cosine scheduler, with a minimum value of $0.04$. The training was configured for 5 epochs with a batch size of 48, using 8 data loading workers, pinned memory, persistent loaders, and a prefetch factor of 2.

\paragraph{\textbf{\textit{Repository contents.}}} The release includes: (i) modular training/inference code for the three core modules, namely \textit{Morphological Analysis}, \textit{Contextual Alignment}, and \textit{Philological grounding}, within which all supporting models (e.g., detection/recognition/sorting for rubbings, glyph-analysis ensemble, contextual restoration, inscription-level interpretation utilities, and historical document retrieval) are implemented; (ii) pretrained weights for all released components; (iii) unified configuration files; (iv) data schemas and parsers for rubbings, transcriptions, and parallel interpretation pairs; (v) evaluation scripts (component-wise and end-to-end), ablation switches, and bootstrapped reporting utilities; (vi) annotation and curation guidelines; and (vii) a model description document, including intended use, limitations, and known failure modes.

\paragraph{\textbf{\textit{Reproducibility.}}} A live demo provides end-to-end reproduction of the AlphaOracle pipeline at  \url{http://vlrlabmonkey.xyz:7685/?lan=en}. For local runs, we include minimal fixed-seed scripts and configs in the repository. 

\paragraph{\textbf{\textit{Data access.}}} For copyrighted or licensed sources (e.g., large-scale rubbings and classical texts), we redistribute only derived indices/metadata and provide data scripts to recreate inputs from publicly available or institution-licensed editions. 
We do not redistribute the original high-resolution images or scans. Our data usage aligns with the principles of fair use as outlined in the Copyright Law of the People's Republic of China. Specifically, our work involves limited use of published materials for research purposes, without affecting the normal exploitation of the original works or prejudicing the legitimate rights of copyright holders. All released materials consist solely of derived resources (e.g., indices, metadata, annotations, and processed representations) generated through our pipeline for non-commercial academic use, and are not intended to substitute for the original sources. Researchers seeking access to the original images should obtain them directly from the corresponding authorized institutions, subject to the terms and conditions of those sources.
The online demo (\url{http://vlrlabmonkey.xyz:7685/?lan=en}) provides an interactive demonstration of the workflow, while additional documentation and usage instructions are available in the GitHub repository \url{https://github.com/Yuliang-Liu/AlphaOracle}.

The framework is released under the Apache license, and we welcome community feedback, issue reports, and contributions via pull requests.

\section{Reading-order parsing for ``Duizhen''}\label{sec:reading_order_parasing}
Oracle bone inscriptions on plastrons often exhibit non-linear spatial organization, including partially symmetric ``Duizhen'' layouts. In our rubbing parsing module, such cases are handled within the same graph-based formulation used for ordinary layouts. Specifically, all detected character instances are connected in a complete graph, and the graph neural network predicts sentence roles and pairwise adjacency relations from character categories and spatial encodings. Therefore, the model does not rely on hand-crafted rules for mirrored or non-linear arrangements, but instead learns reading-order transitions directly from annotated examples.

For characters separated by cracks or visually discontinuous regions, our method does not perform physical fragment rejoining. It operates on rubbings that have already been rejoined and collated in authoritative digitized collections. Accordingly, adjacency prediction is based on the final expert-collated two-dimensional layout of the rubbing.

\begin{figure}[ht!]
    \vspace{-10pt}
    \centering
    \includegraphics[width=0.3\linewidth]{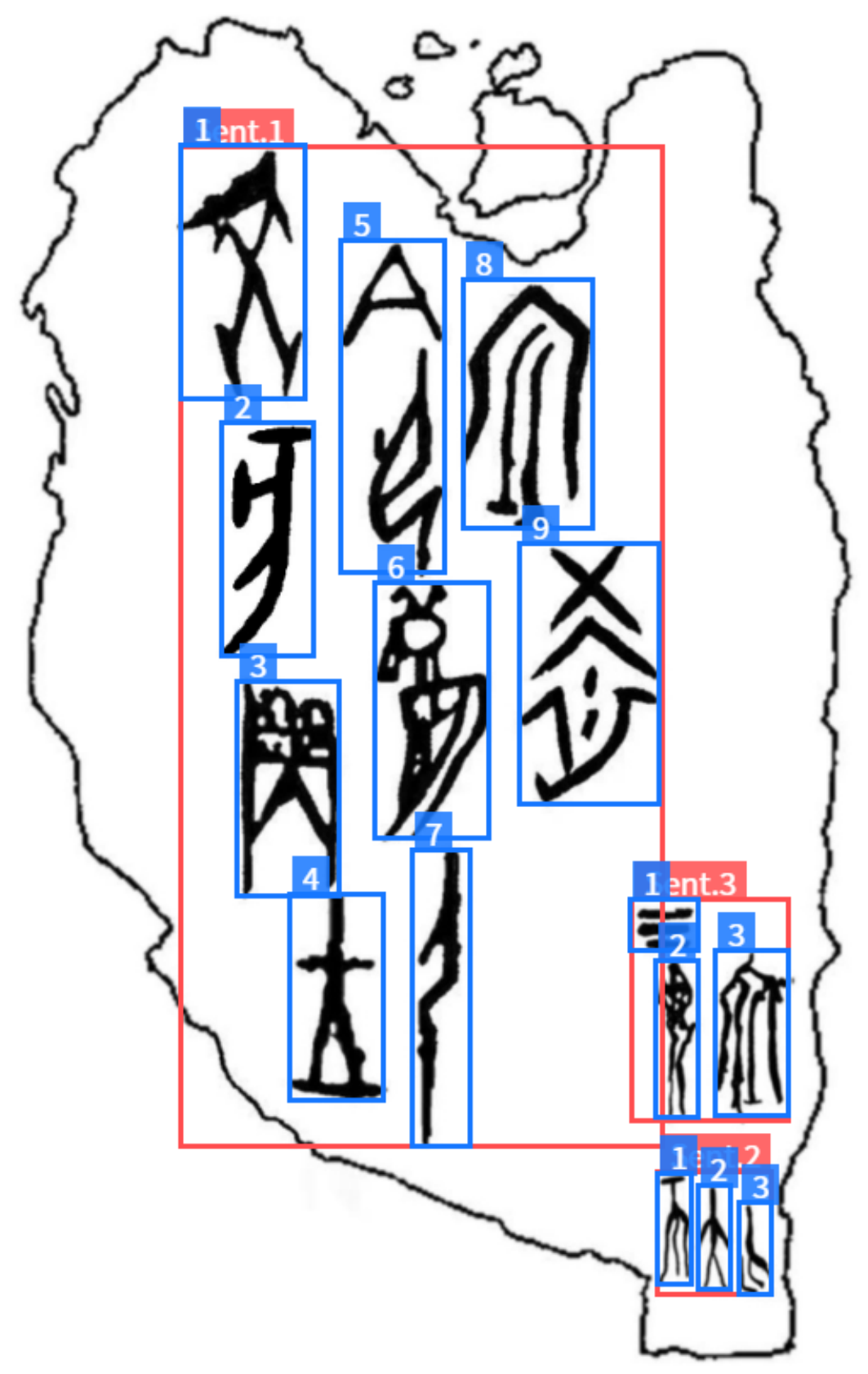}
    \caption{\textbf{Visualization of the reading-order parsing result for ``\texttt{\CJKtrad{屯}} 307''} The rubbing contains three inscriptional sequences, namely (1) ``\CJKtrad{癸亥貞王令龝人宓勞}'', (2) ``\CJKtrad{勿宓三}'', and (3) ``\CJKtrad{不大出}'', where (1) and (2) constitute a typical ``Duizhen'' pattern. Red boxes indicate the predicted sentence-level segmentation, while blue boxes and their indices represent the recovered character order within each sequence. The example shows that the proposed model can correctly distinguish multiple sequences and recover their internal reading order even under partially symmetrical and non-linear layouts.}
    \label{fig:Duizhen}
\end{figure}

Figure \ref{fig:Duizhen} shows a representative ``Duizhen'' example from ``\CJKtrad{屯} 307''. The rubbing contains three inscriptional sequences: (1) ``\CJKtrad{癸亥貞王令龝人宓勞}'', (2) ``\CJKtrad{勿宓三}'', and (3) ``\CJKtrad{不大出}'', where (1) and (2) form a typical ``Duizhen'' pair. Despite the partially symmetric arrangement, the model correctly separates the three sequences and recovers the reading order within each sequence.

\section{User interface}\label{sec:user_interface}

\begin{figure}[!htbp]
\centering

\begin{subfigure}{0.46\linewidth}
    \imagewithbadge{A}{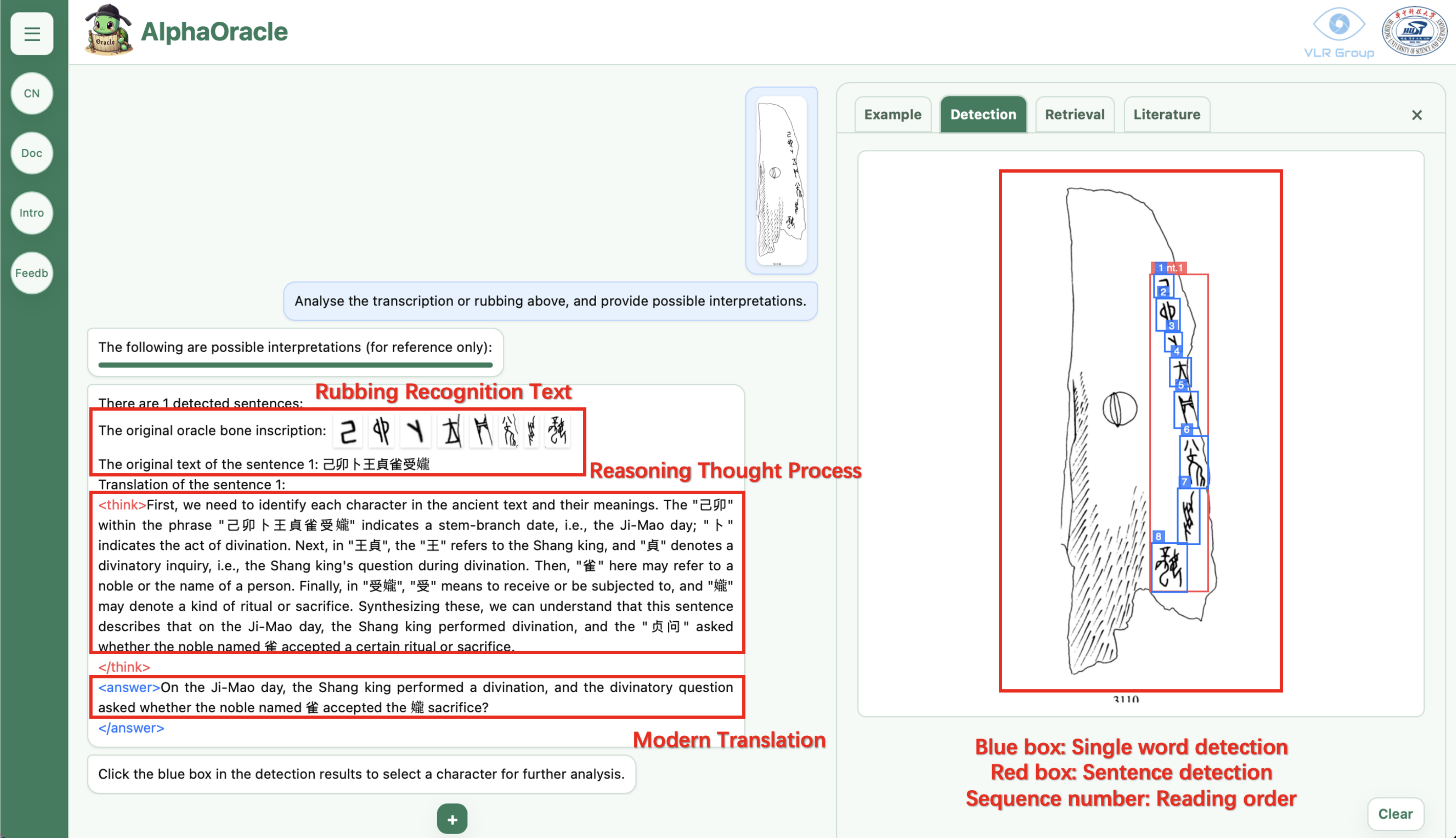}
\end{subfigure}
\begin{subfigure}{0.53\linewidth}
    \imagewithbadge{B}{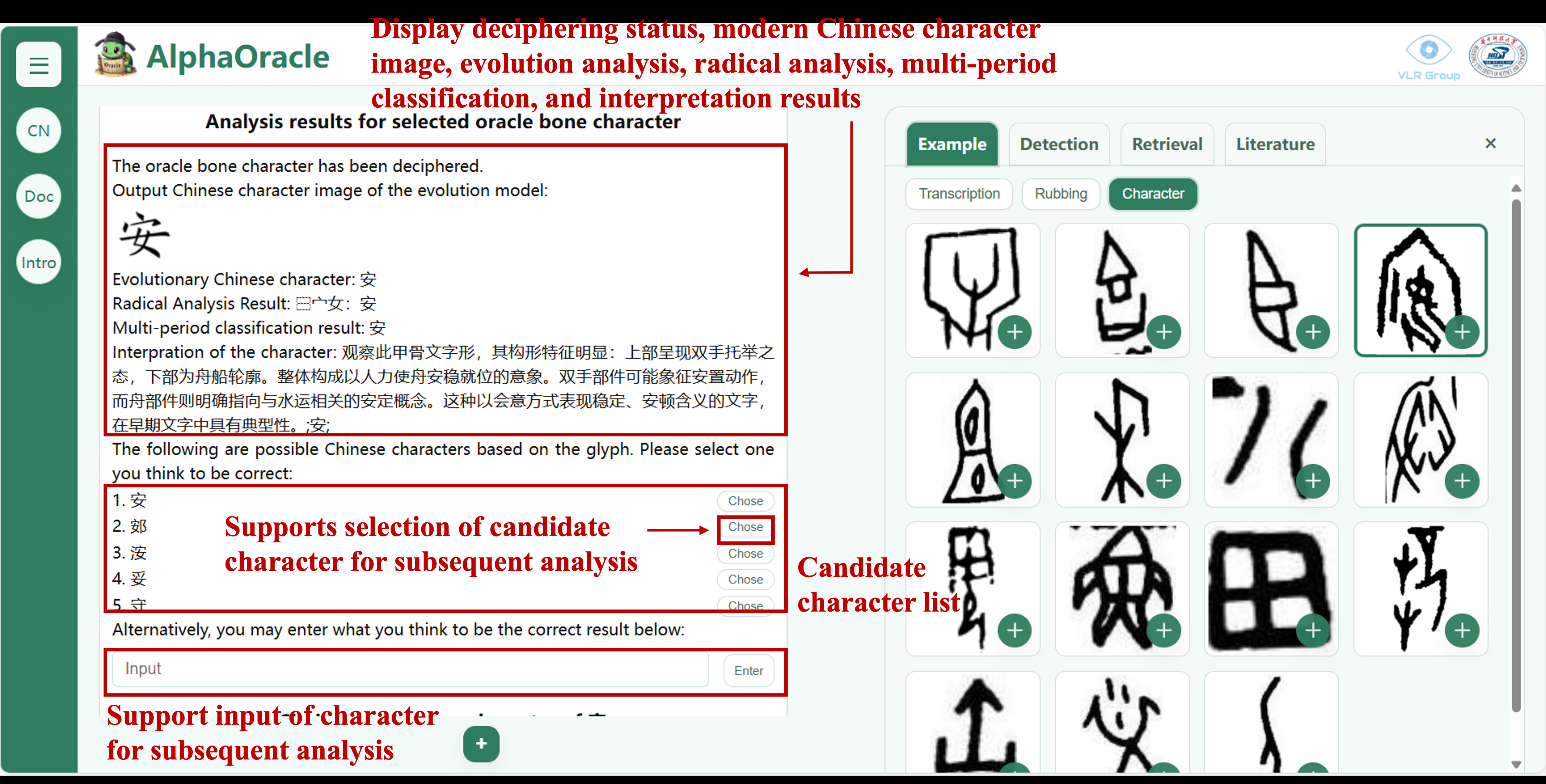}
\end{subfigure}

\begin{subfigure}{0.51\linewidth}
    \imagewithbadge{C}{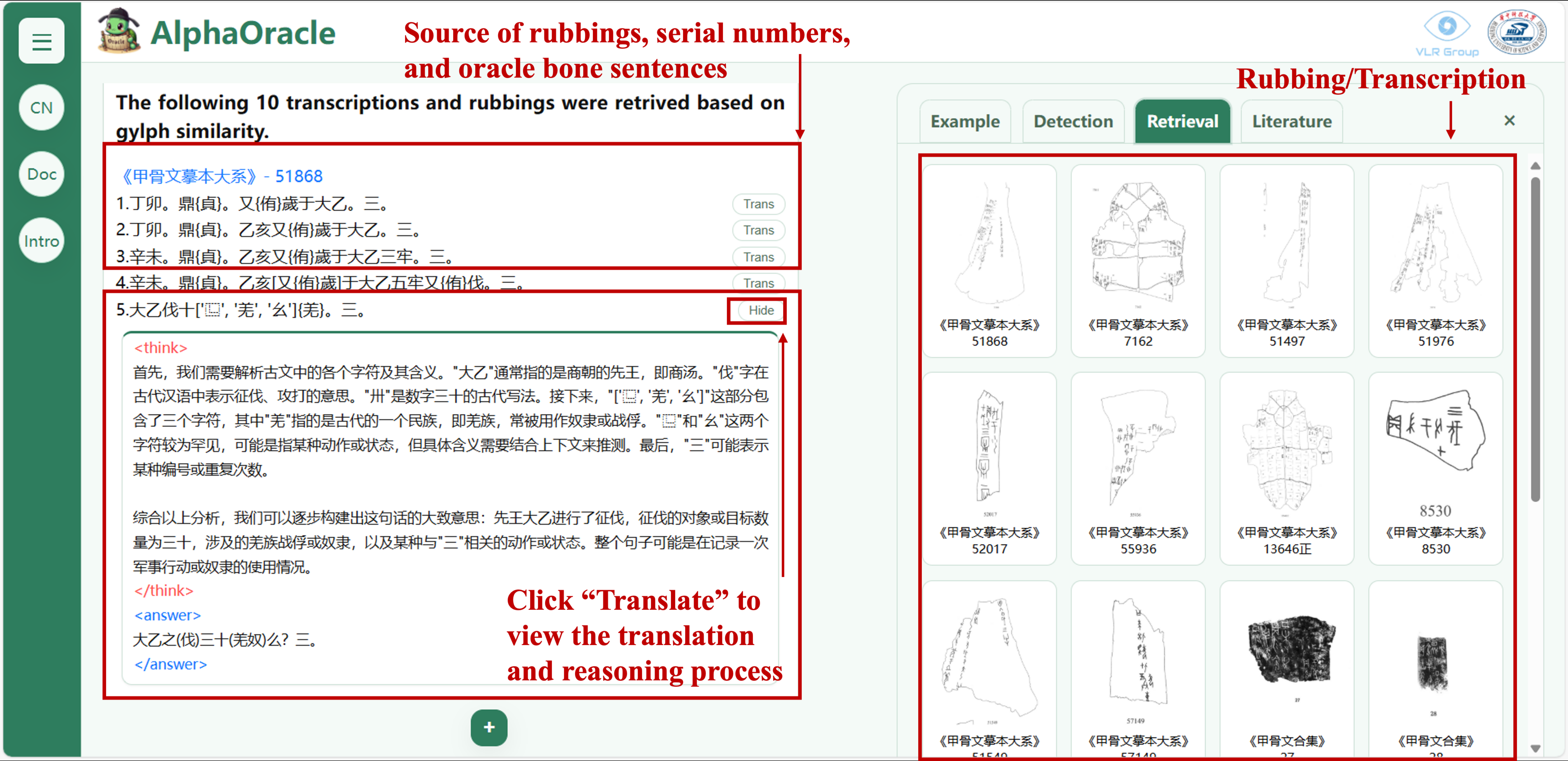}
\end{subfigure}
\begin{subfigure}{0.48\linewidth}
    \imagewithbadge{D}{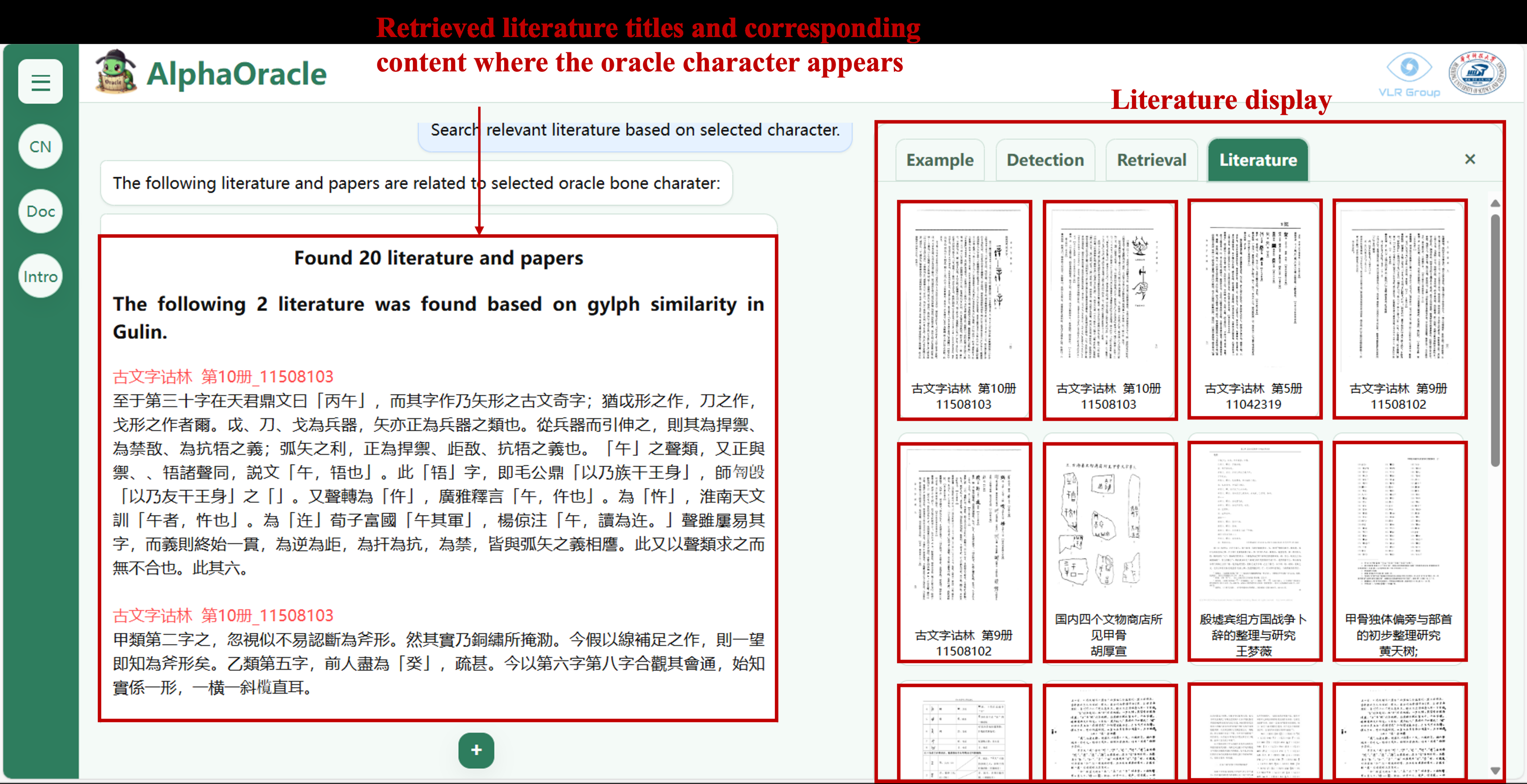}
\end{subfigure}

\caption{\textbf{Interface demonstration of four major functions in AlphaOracle}
(A) \textit{Rubbing Processing}.
(B) \textit{Morphological Analysis}.
(C) \textit{Contextual Alignment}.
(D) \textit{Philological Grounding}.}
\label{fig:ui-demo}

\end{figure}

To make the practical usage of AlphaOracle clearer to readers, we provide representative screenshots of the system interface in Figure~\ref{fig:ui-demo}. Figure~\ref{fig:ui-demo}A shows the \textit{rubbing processing} interface, where the system performs character detection, recognition, sentence segmentation, and intra-sentence ordering on oracle bone rubbings. Users may further select specific sentences or individual characters of interest for downstream analysis. Figure~\ref{fig:ui-demo}B shows the \textit{morphological analysis} interface, where the system presents morphological characteristics and diachronic evolution evidence for a target character. Figure~\ref{fig:ui-demo}C shows the \textit{contextual alignment} interface, where visually similar glyphs are retrieved and displayed in their full rubbing or transcription contexts, allowing researchers to inspect parallel occurrences across inscriptions and view corresponding modern-language translations for comparison. Figure~\ref{fig:ui-demo}D shows the \textit{philological grounding} interface, where the system retrieves relevant discussions from ancient classical texts and modern scholarship based on candidate readings; users can further trace specific literature sources and examine the original analyses. Together, these interfaces illustrate how AlphaOracle supports evidence inspection, cross-source comparison, and scholar-centered interactive exploration during decipherment research, with the results presented as different views of the system’s comprehensive decipherment report. In practice, the online demonstration system provides additional interactive capabilities beyond the automated workflow described in the paper, enabling users to browse retrieved evidence more flexibly, inspect alternative contexts, and explore the system’s analytical outputs.

\section{Limitation}\label{sec:limitation}
Many graphs admit multiple context dependent readings, and some likely functioned as clan names, personal names, or place names that did not persist into later traditions. In addition, many forms are hapax, damaged, or unevenly represented in digitized corpora, and orthographic variation across periods and workshops further increases uncertainty. These factors may bias retrieval and modeling, so modern language outputs should be treated as hypotheses for expert adjudication rather than definitive conclusions. This challenge is especially acute for characters with extremely limited occurrences or severe physical damage, where sparse contextual evidence weakens the reliability of distributional analysis. AlphaOracle partially mitigates this limitation through complementary evidence sources: morphological analysis relies on the visual and structural properties of individual characters rather than frequency alone, allowing rare forms to be compared with deciphered characters through shared components and broader diachronic evolution patterns, while the philological grounding module can retrieve visually similar forms from bronze script, seal script, and later materials to provide additional cross period references. At the same time, AlphaOracle explicitly flags low confidence cases instead of presenting them as definitive readings. Some characters, especially proper nouns whose referents have been lost, may remain resistant to full decipherment by any method. We therefore view AlphaOracle not as a system that resolves all such cases, but as one that organizes partial evidence and supports expert judgment more efficiently.

In addition, the philological grounding module relies on a curated literature base whose reliability depends on the quality and currency of the underlying sources. The current corpus, comprising approximately 23{,}755 publications, books, and reference materials, is primarily drawn from expert-recommended, authoritative, and publicly recognized scholarly sources. AlphaOracle adopts a human-in-the-loop update mechanism, through which outdated or controversial materials can be removed or deprioritized, while newer scholarship can be incorporated over time. Since there is often no universal standard for determining which interpretation is definitively correct in this domain, the system is intended to provide source indexing, evidence aggregation, and comparative summaries across multiple scholarly viewpoints rather than final judgments. This design helps experts review relevant evidence more efficiently, while final interpretive decisions remain with human experts.

Moreover, uneven digitization quality and incomplete corpora dataset bias may affect AlphaOracle in several ways. Characters or inscriptions that are better preserved, more frequently studied, or more heavily represented in digitized corpora may receive stronger contextual support and therefore more confident interpretations, while rarer or less well represented materials may be disadvantaged. In addition, biases in the existing philological literature may shape retrieval results, since some scholarly traditions or interpretive perspectives are more extensively documented than others. Digitization quality may also introduce uneven effects across character types, as forms with fine strokes or severe damage are more vulnerable to recognition errors. These factors may influence both retrieval and modeling, and therefore the system’s outputs should be interpreted cautiously as evidence supported hypotheses rather than definitive conclusions. In addition, this study does not explicitly model the Five-Period division of oracle bone script due to the limited availability of large-scale period-labeled data, and incorporating such fine-grained periodization remains an important direction for future work.

\FloatBarrier

\FloatBarrier

\FloatBarrier
\begin{table}[!htbp]
\centering
\caption{AlphaOracle datasets: processing method, sources and sizes}
\resizebox{1\textwidth}{!}{
\footnotesize
\setlength{\tabcolsep}{6pt}
\renewcommand{\arraystretch}{1.15}

\newlength{\rowgap}
\setlength{\rowgap}{1.10\baselineskip}
\setlength{\extrarowheight}{\rowgap}

\newcolumntype{M}[1]{>{\centering\arraybackslash}m{#1}}
\begin{adjustbox}{max width=\textwidth}
\begin{tabular}{
  M{0.18\linewidth}
  M{0.22\linewidth}
  M{0.34\linewidth}
  M{0.22\linewidth}
}
\toprule
\rowcolor{gray}\makecell{\textbf{Processing}\\[-2pt]\textbf{method}} &
\textbf{Dataset} &
\textbf{Source} &
\textbf{Size} \\
\midrule

\multirow{7}{=}[\dimexpr -0.38\baselineskip - 7\rowgap/2\relax]{\centering\makecell{Open\\ Source}} &
\cellcolor{blue2}\textbf{HUST-OBC} &
\url{https://github.com/Pengjie-W/HUST-OBC} &
140{,}053 images; 77{,}075 deciphered; 62{,}978 undeciphered \\
& \cellcolor{blue2}\textbf{EVOBC} &
\url{https://github.com/RomanticGodVAN/character-Evolution-Dataset} &
13{,}714 categories; 229{,}170 characters \\
& \cellcolor{blue2}\textbf{OBIMD} &
\url{https://www.jgwlbq.org.cn/dt/oracleFragment} &
10{,}077 rubbings/copies; 94{,}652 characters \\
& \cellcolor{blue2}\textbf{Oracle Character Detection Dataset} &
\url{https://jgw.aynu.edu.cn/home/down/detail/index.html?sysid=3} &
9{,}306 rubbings \\
& \cellcolor{blue2}\textbf{Jiaguwen heji Dataset} &
\url{https://jgw.aynu.edu.cn/home/zl/index.html} &
45{,}364 rubbings \\
& \cellcolor{blue2}\textbf{Classical-Modern} &
\url{https://github.com/NiuTrans/Classical-Modern} &
24 books (segmented passages) \\
& \cellcolor{blue2}\textbf{Shuowen Jiezi Dataset} &
\url{https://github.com/lagrangelandada/shuowen} &
7{,}714 character entries (definitions) \\
\midrule
\multirow{6}{=}[\dimexpr -0.30\baselineskip - 6\rowgap/2\relax]{\centering\makecell{Digital\\ Engine}} 
& \cellcolor{blue2}\textbf{Jiaguwen Moben Daxi Dataset} &
\emph{Book: Jiaguwen Moben Daxi} &
73{,}883 transcriptions \\
& \cellcolor{blue2}\textbf{Jiaguwen Jiaoshi Zongji Dataset} &
\emph{Book: Jiaguwen Jiaoshi Zongji} &
45{,}413 rubbings \\
& \cellcolor{blue2}\textbf{Jiaguwen Jingcui Shiyi Dataset} &
\emph{Book: Jiaguwen Jingcui Shiyi} &
1500 OBIs with vernacular explanations \\
& \cellcolor{blue2}\textbf{Oracle-Bone Studies Papers Dataset} &
\url{https://jgw.aynu.edu.cn/home/wx/index.html} &
23{,}755 papers \\
& \cellcolor{blue2}\textbf{Guwenzi Gulin Dataset} &
\emph{Book: Guwenzi Gulin} &
11 volumes; 384{,}542 character images \\
& \cellcolor{blue2}\textbf{Jiaguwenzi Gulin Dataset} &
\emph{Book: Jiaguwenzi Gulin} &
1{,}889 oracle-bone entries (definitions) \\

\bottomrule
\end{tabular}
\end{adjustbox}
}
\label{tab:alphaoracle-datasets}
\end{table}

\newpage

\putbib
\end{bibunit}

\end{document}